\UseRawInputEncoding
\documentclass[journal]{new-aiaa} 
\usepackage[utf8]{inputenc}

\usepackage{graphicx}
\usepackage{float}
\usepackage{amsmath}
\usepackage[version=4]{mhchem}
\usepackage{siunitx}
\usepackage{longtable,tabularx}
\usepackage{bm}
\usepackage{siunitx}
\usepackage{caption}
\usepackage{subcaption}

\title{A Machine Learning Approach to Classify Vortex Wakes of Energy Harvesting Oscillating Foils}

\author{Bernardo Luiz R. Ribeiro\footnote{Graduate Research Assistant, Department of Engineering Physics, AIAA Student Member.}}
\author{Jennifer A. Franck\footnote{Assistant Professor, Department of Engineering Physics, AIAA Senior Member.}}
\affil{University of Wisconsin---Madison, Madison, WI, 53706}

\begin{document}

\maketitle

\textbf{A machine learning model is developed to establish wake patterns behind oscillating foils whose kinematics are within the energy harvesting regime. The role of wake structure is particularly important for array deployments of oscillating foils, since the unsteady wake highly influences performance of downstream foils. This work explores \bm{$46$} oscillating foil kinematics, with the goal of parameterizing the wake based on the input kinematic variables and grouping vortex wakes through image analysis of vorticity fields. A combination of a convolutional neural network (CNN) with long short-term memory (LSTM) units is developed to classify the wakes into three classes. To fully verify the physical wake differences among foil kinematics, a convolutional autoencoder combined with k-means++ clustering is utilized and four different wake patterns are found. With the classification model, these patterns are associated with a range of foil kinematics. Future work can use these correlations to predict the performance of foils placed in the wake and build optimal foil arrangements for tidal energy harvesting.}
\footnote[3]{Presented at the AIAA Aviation 2021 Forum, August 2-6, 2021. Virtual Event. AIAA 2021-2947}

\section{Introduction} \label{intro}

This paper utilizes a machine learning approach to classify vortex wake structures behind an energy harvesting oscillating foil. Understanding the vortex formation and resulting wake structure reveals information about the upstream disturbance, which could potentially be linked to the underlying flow conditions and/or oscillating foil kinematics. Although commonly used in propulsive applications, oscillating foils can also extract energy from the flow in a similar manner as a rotational turbine \cite{Young2014, Xiao2014}. Furthermore, due to the coherent vortex wake of opposite vortex signs generated from the upstroke and downstroke foil motion there is a potential for cooperative motion within tightly packed array configurations to improve performance \cite{oshkaidumas2022}. In order to create control laws to optimize performance in array configurations, it is critical to fully understand and model the wake structure as a function of flow conditions and foil kinematics. 

Traditional wake structure characterization is commonly investigated for cylindrical bluff bodies, such as the canonical work of Williamson and Roshko \cite{williamson1988} who described vortex wakes with a `\textit{mS + nP}' notation, where \textit{m} is the number of single vortices (\textit{S}) shed per cycle, and \textit{n} is the number of clockwise/counter-clockwise vortex pairs (\textit{P}). This notation has been propagated to oscillating foils with some success when the foil is in pure pitching \cite{schnipper2009, koochesfahani2012} or plunging \cite{lai1999} motion. However, previous investigations focused on kinematics for propulsive foils and did not consider the wakes generated by an oscillating foil in the energy harvesting regime. When used for energy harvesting, the foil kinematics are characterized by a lower non-dimensional frequency and higher pitch/heave amplitudes compared to oscillating foil propulsion. The high amplitudes result in a rich variety of wakes with multiple vortices shed each foil stroke that are often chaotic and not easily identified by the `\textit{mS + nP}' notation \cite{RibeiroFranck2020, RibeiroFranck2021}.

To assist in the wake modeling of bluff bodies outside the canonical characterization, various machine learning techniques can be utilized. Particularly, convolutional neural networks (CNN) receive much interest due to the ability to process data from images for pattern recognition and prediction \cite{brunton2020}. For instance, a CNN is implemented to analyze cylinder and airfoil flow for various Reynolds numbers and obtained accurate predictions of the velocity field \cite{lee2019, bhatnagar2019} and force \cite{morimoto2021} when compared to numerical data. Another example of a CNN is for vortex identification procedure \cite{deng2019} which does not require user-input for thresholding such as \textit{Q}-criterion \cite{Hunt1988} or $\lambda_2$ criterion \cite{jeong1995}.

When analyzing unsteady flows, the time evolution of structures can be captured with recurrent neural networks with the use of long short-term memory (LSTM) \cite{hochreiter1997}, which can predict flow quantities by holding information from an input sequence, and not simply from a single input \cite{han2019}. Using flow information from the past five timesteps, Nakamura et al. \cite{nakamura2021} predicted turbulent structures in a channel flow using a convolutional autoencoder combined with LSTM. LSTM was also implemented in a dynamic wind farm wake model that predicts the main features of unsteady wind turbine wakes almost as well as high-fidelity computational models \cite{zhang2020}. The integration of convolutional layers and LSTM units has also predicted unsteady flow fields behind bluff bodies such as a cylinder and a foil \cite{han2019}.

The goal of this investigation is to develop a neural network that integrates convolutional layers and LSTM units towards classification of the bluff body wake structures behind oscillating foils. Using such classification will enable predictive models for these chaotic wakes, and inform performance optimization of arrays of foils operating as energy harvesters. Classification models have been previously implemented for wakes of propulsive foils \cite{colvert2018, pollard2021} and cylinders \cite{li2020}. This work classifies and clusters the drag-based wake from an oscillating foil in energy harvesting mode, which distinguishes itself from propulsive oscillating foils due to the lower oscillating frequency, higher pitch, and higher heave amplitudes \cite{godoy-diana2008}. Prior work has also largely classified wake structures from point measurements within the wake. In contrast, this research classifies vortex wake structures from vorticity flow fields. Using images as the input data allows for connections linking wake features such as the size and strength of vortices with the underlying foil kinematics. Due to the unsteady interactions between vortices that affect the wake trajectory behind each foil configuration, the LSTM network aims to provide information on how each wake image is linked throughout time, improving the classification outcomes. The results of the supervised classification groupings are compared directly against an unsupervised convolutional autoencoder (CAE) and clustering methodology \cite{calvet2020} to confirm and update the boundaries between classes. Finally, the physics of the wake structure in each of the classified groups are explained and correlated with energy harvesting efficiency and foil kinematics.

Section \ref{s:CFD} gives an overview of the foil kinematics and simulation methods followed by the initial groupings for the classification model in Section \ref{s:vortexstrength}. An overview of the classification network architecture and results are presented in Section \ref{s:classification}, and Section \ref{s:clustering} describes the groupings, and updates obtained from an unsupervised clustering of the input images with conclusions presented in Section \ref{s:conclusion}.

\section{Computational Methods} \label{s:CFD}

This section discusses the computational fluid dynamics methods, including the foil kinematics, flow solver and mesh.

\subsection{Foil Parameters}

The foil kinematics are defined by three parameters, namely pitch amplitude, $\theta_o$, heave amplitude, $h_o$, and reduced frequency, $fc/U_\infty$, where $c$ is the foil's chord length and $U_\infty$, the freestream velocity. To generate a range of energy harvesting vortex wake structures, $46$ unique kinematics are prescribed to a $10\%$ thick elliptical foil through numerical simulations. The foil motion is sinusoidal in pitch and heave, utilizing a range of frequencies and amplitudes previously established as effective at energy harvesting \cite{kindum2008, RibeiroFranck2020}. As opposed to an airfoil geometry, a thin elliptical foil is utilized as geometry has shown to have minor effects on efficiency \cite{Kim2017} and the fore-aft symmetry is desirable for harvesting energy from tidal flows. The kinematics with respect to time $t$ are described in lab-fixed coordinates as

\begin{equation}
h(t)=-h_{o}\cos(2\pi f t) 
\label{eq:heave}
\end{equation}

\noindent and 

\begin{equation}
\theta(t)=-\theta_{o}\cos(2\pi f t + \pi/2),
\label{eq:pitch}
\end{equation}

\noindent where $h(t)$ and $\theta(t)$ are the prescribed heave and pitch motions, respectively, with pitching about the center-chord. The phase difference between pitch and heave is fixed at $\pi/2$, which is found to yield the optimal energy harvesting performance \cite{zhu2011}. At $t=0$ the foil is at the bottom of its heave stroke. Heave and pitch are changing simultaneously during foil motion, which creates a time-varying relative angle of attack with respect to the freestream flow given by

\begin{equation}
\alpha_{rel}(t) = \tan^{-1} (-\dot{h}(t)/U_{\infty}) + \theta(t),
\label{eq:alpharel}
\end{equation} 

\noindent with $\dot{h}(t)$ representing the time derivative of the heave motion. A characteristic relative angle of attack is evaluated when the foil is at maximum pitch and maximum heave velocity, which occurs at one quarter of the cycle period $T$, or

\begin{equation}
\alpha_{T/4} =\alpha_{rel}(t/T = 0.25) = \tan^{-1} (-{2\pi f h_0}/U_{\infty}) + \theta_0.
\label{eq:at4}
\end{equation}

\begin{figure}[H]
\centering
\includegraphics[width=0.65\textwidth]{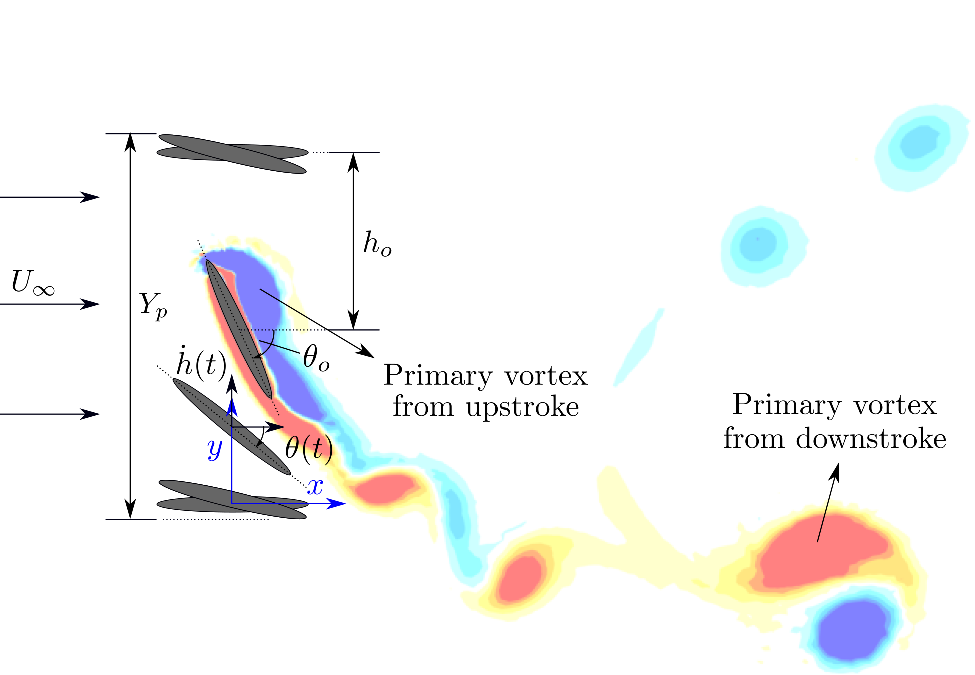}
\caption{Foil kinematics: pitch amplitude (\bm{$\theta_o$}), heave amplitude (\bm{$h_o$}) and foil's swept area (\bm{$Y_p$}) shown. Spanwise vorticity at non-dimensional time \bm{$t/T=0.50$} for the kinematics \bm{$fc/U_\infty=0.10; h_o=1.00; \theta_o=75^{\circ}$} is displayed as an example of the oscillating foil wake.}
\label{f:singlefoil}
\end{figure}

The foil motion along with the foil parameters are illustrated in Figure \ref{f:singlefoil}. The primary vortex, or the first vortex formed on the suction surface during each half stroke, is highlighted for the case $fc/U_\infty=0.10; h_o=1.00; \theta_o=75^{\circ}$. Previous vortices originating from prior oscillating cycles provide an overall view of the wake.

To evaluate performance the energy harvesting efficiency is defined as

\begin{equation}
\eta = \frac{\bar{P}}{\frac{1}{2}\rho U_{\infty}^3 Y_p},
\label{eq:eta}
\end{equation}

\noindent which is the ratio of the average power extracted, $\bar{P}$, to the power available in the freestream velocity throughout the swept area $Y_p$. The power extracted is defined as

\begin{equation}
P(t)=F_y \dot{h} + M_z \dot{\theta},
\label{eq:power}
\end{equation}

\noindent where $F_y$ and $M_z$ are the vertical force and spanwise moment on the foil, respectively. To remove small cycle-to-cycle variations the efficiency is phase-averaged over the last three cycles of simulation.

A wide range of kinematics within the energy harvesting regime is considered, all of which directly influence $\alpha_{T/4}$. Table \ref{t:kin} outlines all simulated foil kinematics, with seven sets of different kinematics chosen specifically with the same $\alpha_{T/4}$. All quantities reported in this manuscript are non-dimensionalized by the freestream velocity, $U_\infty$, and $c$ and thus, $f^*=fc/U_\infty$ and $h_o^*=h_o/c$. Reduced frequency, pitch amplitude and heave amplitude are varied from $f^* = 0.10 - 0.15$, $ \theta_o = 40^{\circ} - 80^{\circ}$, $h_o^* = 0.50 - 1.50$.

\begin{table}[htbp]
\setlength{\tabcolsep}{8pt}
\centering
\begin{tabular}{rc|rc}
\hline
\textbf{Kinematics} & \bm{$\alpha_{T/4}$} & \textbf{Kinematics} & \bm{$\alpha_{T/4}$} \\ \hline
$^{(*,+)}f^* = 0.15 ; h_o^* = 1.25 ; \theta_o = 55^{\circ}$ & $5.3^{\circ}$ &
$f^* = 0.12 ; h_o^* = 0.50 ; \theta_o = 50^{\circ}$ & $29.3^{\circ}$ \\
$^{(*,+)}f^* = 0.12 ; h_o^* = 1.50 ; \theta_o = 55^{\circ}$ & $6.5^{\circ}$ &
$f^* = 0.10 ; h_o^* = 0.75 ; \theta_o = 55^{\circ}$ & $29.8^{\circ}$ \\
$^{(*,+)}f^* = 0.15 ; h_o^* = 1.00 ; \theta_o = 50^{\circ}$ & $6.7^{\circ}$ &
$^{(*,+)}f^* = 0.12 ; h_o^* = 0.75 ; \theta_o = 60^{\circ}$ & $30.5^{\circ}$ \\
$^{(+)}f^* = 0.10 ; h_o^* = 1.00 ; \theta_o = 40^{\circ}$ & $7.9^{\circ}$ &
$^{(*,+)}f^* = 0.15 ; h_o^* = 1.00 ; \theta_o = 75^{\circ}$ & $31.7^{\circ}$ \\
$^{(*,+)}f^* = 0.12 ; h_o^* = 1.00 ; \theta_o = 45^{\circ}$ & $8.0^{\circ}$ &
$^{(*,+)}f^* = 0.12 ; h_o^* = 1.25 ; \theta_o = 75^{\circ}$ & $31.7^{\circ}$ \\
$f^* = 0.15 ; h_o^* = 0.75 ; \theta_o = 45^{\circ}$ & $9.8^{\circ}$ &
$f^* = 0.10 ; h_o^* = 0.50 ; \theta_o = 50^{\circ}$ & $32.6^{\circ}$ \\
$^{(*,+)}f^* = 0.12 ; h_o^* = 0.75 ; \theta_o = 40^{\circ}$ & $10.5^{\circ}$ &
$^{(*,+)}f^* = 0.10 ; h_o^* = 1.00 ; \theta_o = 65^{\circ}$ & $32.9^{\circ}$ \\
$^{(*,+)}f^* = 0.15 ; h_o^* = 1.00 ; \theta_o = 55^{\circ}$ & $11.7^{\circ}$ &
$f^* = 0.12 ; h_o^* = 0.50 ; \theta_o = 55^{\circ}$ & $34.3^{\circ}$ \\
$^{(*,+)}f^* = 0.10 ; h_o^* = 1.00 ; \theta_o = 45^{\circ}$ & $12.9^{\circ}$ &
$f^* = 0.10 ; h_o^* = 0.75 ; \theta_o = 60^{\circ}$ & $34.8^{\circ}$ \\
$^{(+)}f^* = 0.12 ; h_o^* = 1.00 ; \theta_o = 50^{\circ}$ & $13.0^{\circ}$ &
$^{(*,+)}f^* = 0.12 ; h_o^* = 0.75 ; \theta_o = 65^{\circ}$ & $35.5^{\circ}$ \\
$^{(*,+)}f^*= 0.10 ; h_o^* = 0.75 ; \theta_o = 40^{\circ}$ & $14.8^{\circ}$ &
$f^* = 0.15 ; h_o^* = 1.00 ; \theta_o = 80^{\circ}$ & $36.7^{\circ}$ \\
$^{(*,+)}f^* = 0.15 ; h_o^* = 1.00 ; \theta_o = 60^{\circ}$ & $16.7^{\circ}$ &
$f^* = 0.10 ; h_o^* = 0.50 ; \theta_o = 55^{\circ}$ & $37.6^{\circ}$ \\
$^{(+)}f^* = 0.12 ; h_o^* = 1.00 ; \theta_o = 55^{\circ}$ & $18.0^{\circ}$ &
$^{(*,+)}f^* = 0.12 ; h_o^* = 1.00 ; \theta_o = 75^{\circ}$ & $38.0^{\circ}$ \\
$f^* = 0.12 ; h_o^* = 0.50 ; \theta_o = 40^{\circ}$ & $19.3^{\circ}$ &
$f^* = 0.12 ; h_o^* = 0.50 ; \theta_o = 60^{\circ}$ & $39.3^{\circ}$ \\
$^{(*,+)}f^* = 0.10 ; h_o^* = 0.75 ; \theta_o = 45^{\circ}$ & $19.8^{\circ}$ &
$f^* = 0.12 ; h_o^* = 0.75 ; \theta_o = 70^{\circ}$ & $40.5^{\circ}$ \\
$^{(+)}f^* = 0.12 ; h_o^* = 0.75 ; \theta_o = 50^{\circ}$ & $20.5^{\circ}$ &
$f^* = 0.10 ; h_o^* = 1.25 ; \theta_o = 80^{\circ}$ & $41.9^{\circ}$ \\
$^{(*,+)}f^* = 0.12 ; h_o^* = 1.25 ; \theta_o = 65^{\circ}$ & $21.7^{\circ}$ &
$^{(*,+)}f^* = 0.10 ; h_o^* = 1.00 ; \theta_o = 75^{\circ}$ & $42.9^{\circ}$ \\
$^{(*,+)}f^* = 0.10 ; h_o^* = 1.25 ; \theta_o = 60^{\circ}$ & $21.9^{\circ}$ & $f^* = 0.12 ; h_o^*= 1.00 ; \theta_o = 80^{\circ}$ & $43.0^{\circ}$ \\
$f^* = 0.12 ; h_o^* = 1.00 ; \theta_o = 60^{\circ}$ & $23.0^{\circ}$ & $f^* = 0.10 ; h_o^* = 0.75 ; \theta_o = 70^{\circ}$ & $44.8^{\circ}$ \\
$f^* = 0.12 ; h_o^* = 0.50 ; \theta_o = 45^{\circ}$ & $24.3^{\circ}$ & $^{(*,+)}f^* = 0.10 ; h_o^* = 0.50 ; \theta_o = 65^{\circ}$ & $47.6^{\circ}$ \\
$^{(*,+)}f^* = 0.15 ; h_o^* = 1.25 ; \theta_o = 75^{\circ}$ & $25.3^{\circ}$ & $f^* = 0.10 ; h_o^* = 1.00 ; \theta_o = 80^{\circ}$ & $47.9^{\circ}$ \\
$^{(*,+)}f^* = 0.15 ; h_o^* = 1.00 ; \theta_o = 70^{\circ}$ & $26.7^{\circ}$ & $f^* = 0.12 ; h_o^* = 0.50 ; \theta_o = 70^{\circ}$ & $49.3^{\circ}$ \\
$^{(+)}f^* = 0.12 ; h_o^* = 1.00 ; \theta_o = 65^{\circ}$ & $28.0^{\circ}$ &
$f^* = 0.12 ; h_o^* = 0.75 ; \theta_o = 80^{\circ}$ & $50.5^{\circ}$ \\
\hline 
\multicolumn{3}{l}{\footnotesize{$^+$ Training data, $^{*,+}$ First cycle in validation data and remaining two cycles are used in the training data}}\\
\end{tabular}
\caption{Summary of all kinematics with their computed \bm{$\alpha_{T/4}$} values. Footnote refers to the unsupervised clustering model.}
\label{t:kin}
\end{table}

\clearpage

\subsection{Flow Solver and Mesh}

The computations utilize an incompressible Navier-Stokes solver performed using a second-order accurate finite volume, pressure-implicit split-operator (PISO) method implemented in $\textit{OpenFOAM}$ \cite{weller1998}. The Reynolds number of $Re_c=1000$ is selected for all simulations to enable a broad sweep of 46 kinematics within a tractable computational time. Prior work comparing experiments with low and high Reynolds number simulations has demonstrated only minor differences between the power generation and wake characteristics across a Reynolds number regime of $1000-50,000$ \cite{RibeiroFranck2020, RibeiroFranck2021}.

\begin{figure}[H]
\centering
\includegraphics[width=0.8\textwidth]{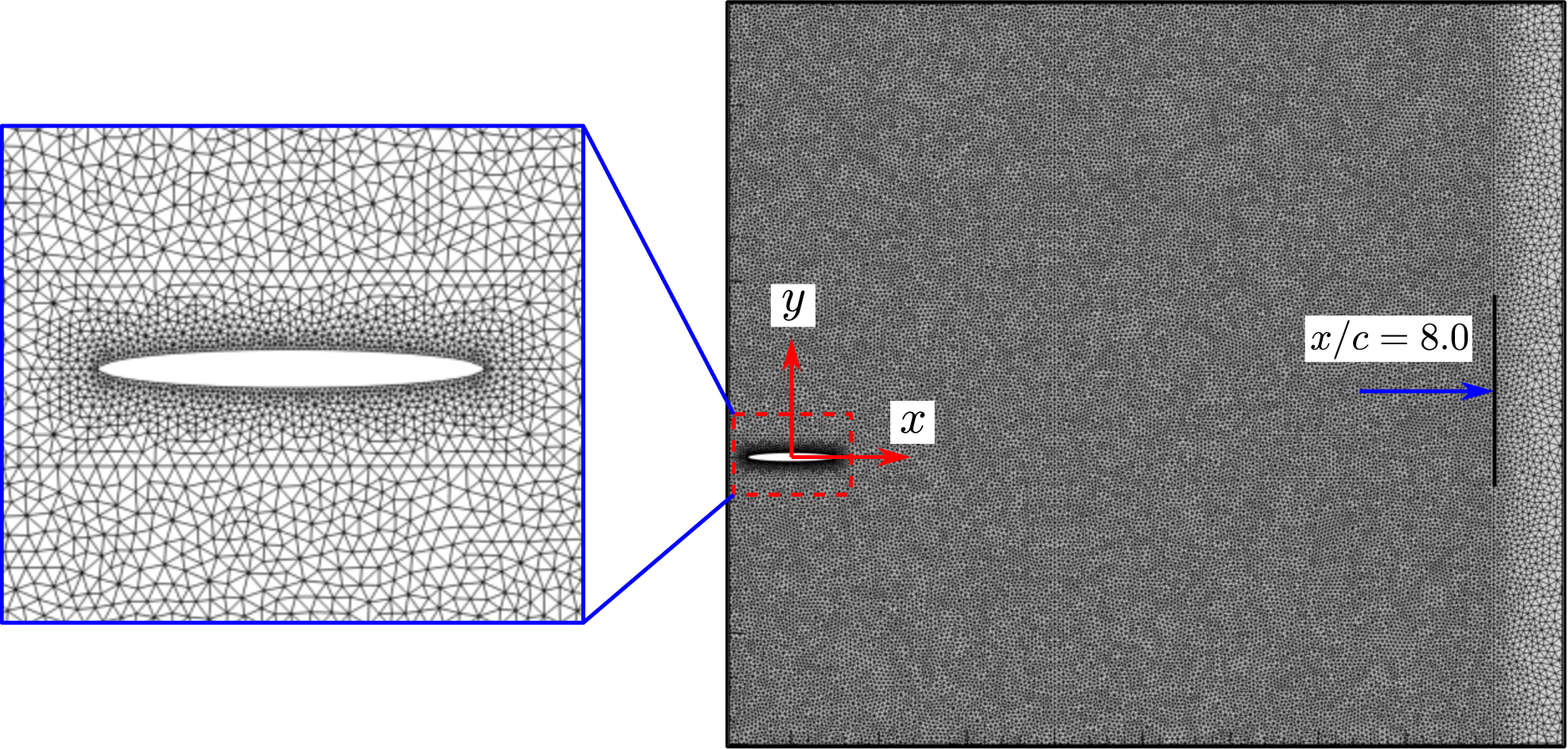}
\caption{Computational domain zoomed in on the foil mesh. The mesh 3A is displayed with its characteristics outlined in Table \ref{t:table_mesh}.}
\label{f:mesh}
\end{figure}

All simulations are performed with a 2D unstructured mesh, with foil motion generated through the boundary condition of a dynamic mesh solver that updates the position of all nodes in the domain at every timestep. The dynamic mesh algorithm utilized in this manuscript is previously validated against a stationary mesh in the work from Ribeiro et al. \cite{RibeiroFranck2021}. The total domain size is $106c$ in the horizontal direction and $100c$ in the vertical direction, with the foil located $50c$ upstream in the $x$ direction and vertically centered when the foil is at the bottom of its heave stroke ($t/T=0$). The mesh is generated using Gmsh \cite{gmsh}, with a subset of the mesh displayed in Figure \ref{f:mesh}.

The boundary conditions entail a no-slip condition at the foil surface with zero pressure gradient, inlet flow on the left boundary, and outlet flow on the top, bottom and right boundaries. Simulations are run for a total of six oscillation cycles, and become stationary after three cycles.

Mesh refinement is evaluated through eight meshes with varying resolution near-foil and in the wake, where $N$ corresponds to the total number of nodes. All mesh characteristics are displayed in Table \ref{t:table_mesh}, where the characteristic wake resolution is measured at $x/c=3.0;y/c=0$. This resolution is held approximately constant until $8c$ downstream as displayed in Figure \ref{f:mesh}. Along the foil surface, a characteristic $\Delta x$ is measured along the mid-chord, and $N_\theta$ represents the total number of nodes on the body in the azimuthal direction. The CPU time is calculated for one oscillation cycle using a single processor on the Intel Cascade architecture.

\begin{table}[htbp]
\centering
\caption{Mesh characteristics.}
\setlength{\tabcolsep}{6pt}
\begin{tabular}{ccccccc}
\hline
& & \textbf{Wake} & & \textbf{Foil} &\\
\textbf{Mesh}   & \bm{$N$} & \bm{$\Delta x$} & \bm{$N_\theta$} & \bm{$\Delta x$} & \bm{$\eta$} & \textbf{CPU time (hrs)}\\
\hline
Mesh 1A & $0.30 \times 10^5$ & 0.14 & 150 & 0.013 & 29.3\% & 1.6 \\
Mesh 2A & $0.48 \times 10^5$ & 0.10 & 150 & 0.013 & 29.7\% & 1.8 \\
Mesh 3A & $1.64 \times 10^5$ & 0.05 & 150 & 0.013 & 26.9\% & 12.8 \\
Mesh 4A & $3.66 \times 10^5$ & 0.03 & 150 & 0.013 & 25.7\% & 41.2 \\ \hline
Mesh 1B & $0.31 \times 10^5$ & 0.14 & 240 & 0.008 & 30.7\% & 1.5 \\
Mesh 2B & $0.49 \times 10^5$ & 0.10 & 240 & 0.008 & 29.9\% & 2.1 \\
Mesh 3B & $1.56 \times 10^5$ & 0.05 & 240 & 0.008 & 26.9\% & 11.3 \\
Mesh 4B & $3.68 \times 10^5$ & 0.03 & 240 & 0.008 & 25.4\% & 40.0 \\ \hline
\end{tabular}
\label{t:table_mesh}
\end{table}

\begin{figure}[H]
\centering
\includegraphics[width=0.84\textwidth]{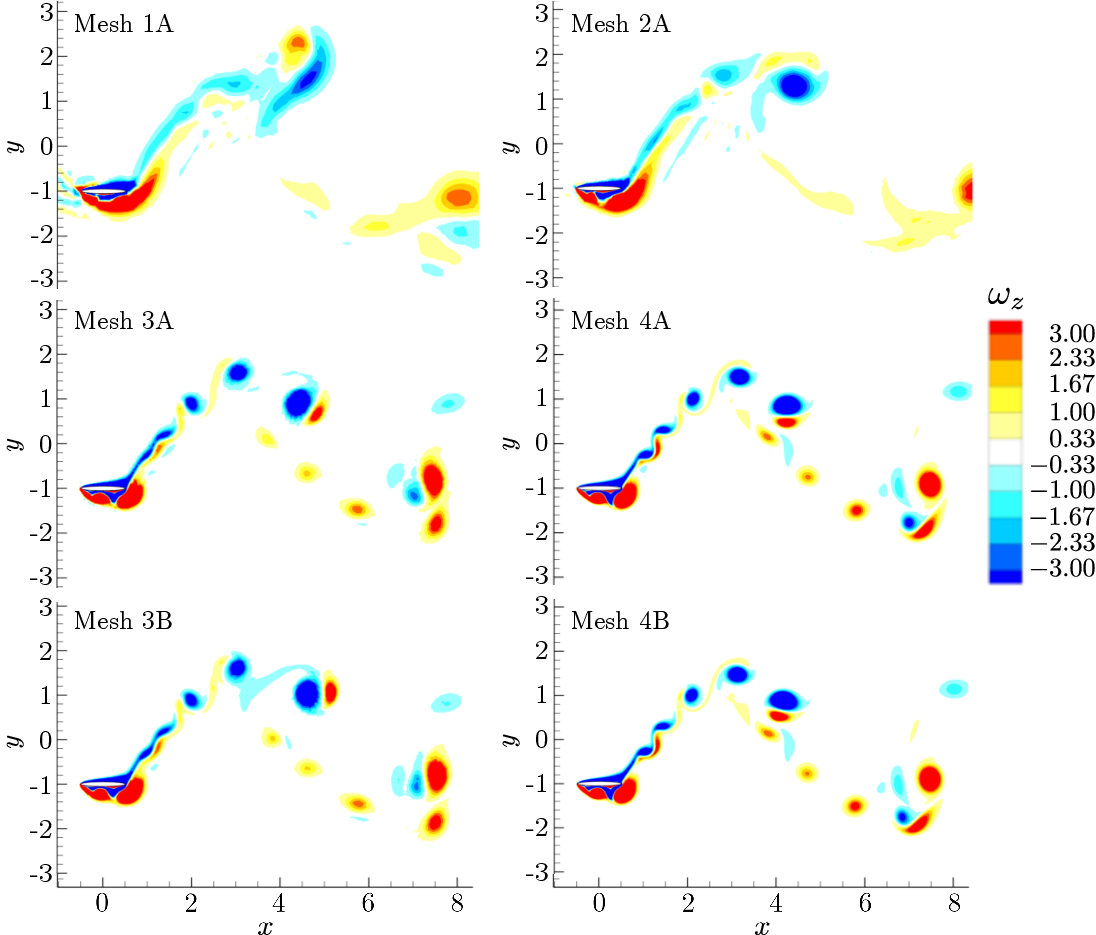}
\caption{Instantaneous vorticity field in the \bm{$z$} direction, \bm{$\omega_z$}. Kinematics: \bm{$f^* = 0.12; h_o^* = 1.00; \theta_o=65^{\circ}$}.}
\label{f:vortmesh}
\end{figure}

Foil efficiency and spanwise vorticity flow fields are evaluated across all meshes. Foil efficiency decreases with increasing wake resolution, demonstrating a $\Delta \eta$ of 3\% between mesh 2 and mesh 3 which has twice the wake resolution. As shown in the vorticity fields, the same number of coherent vortices are observed between meshes 3 and 4, with only small variations in vortex positioning in the wake as displayed in Figure \ref{f:vortmesh}. There is little difference between meshes A and B, indicating both have sufficient near-foil resolution. Therefore, as a balance in computational cost and accuracy of both the foil forces and flow fields, mesh 3A is utilized in all simulations.

\section{Initial Class Selection} \label{s:vortexstrength}

\begin{figure}[H]
\centering
	\begin{subfigure}[b]{0.49\textwidth}
	\centering
        \includegraphics[width=1\linewidth]{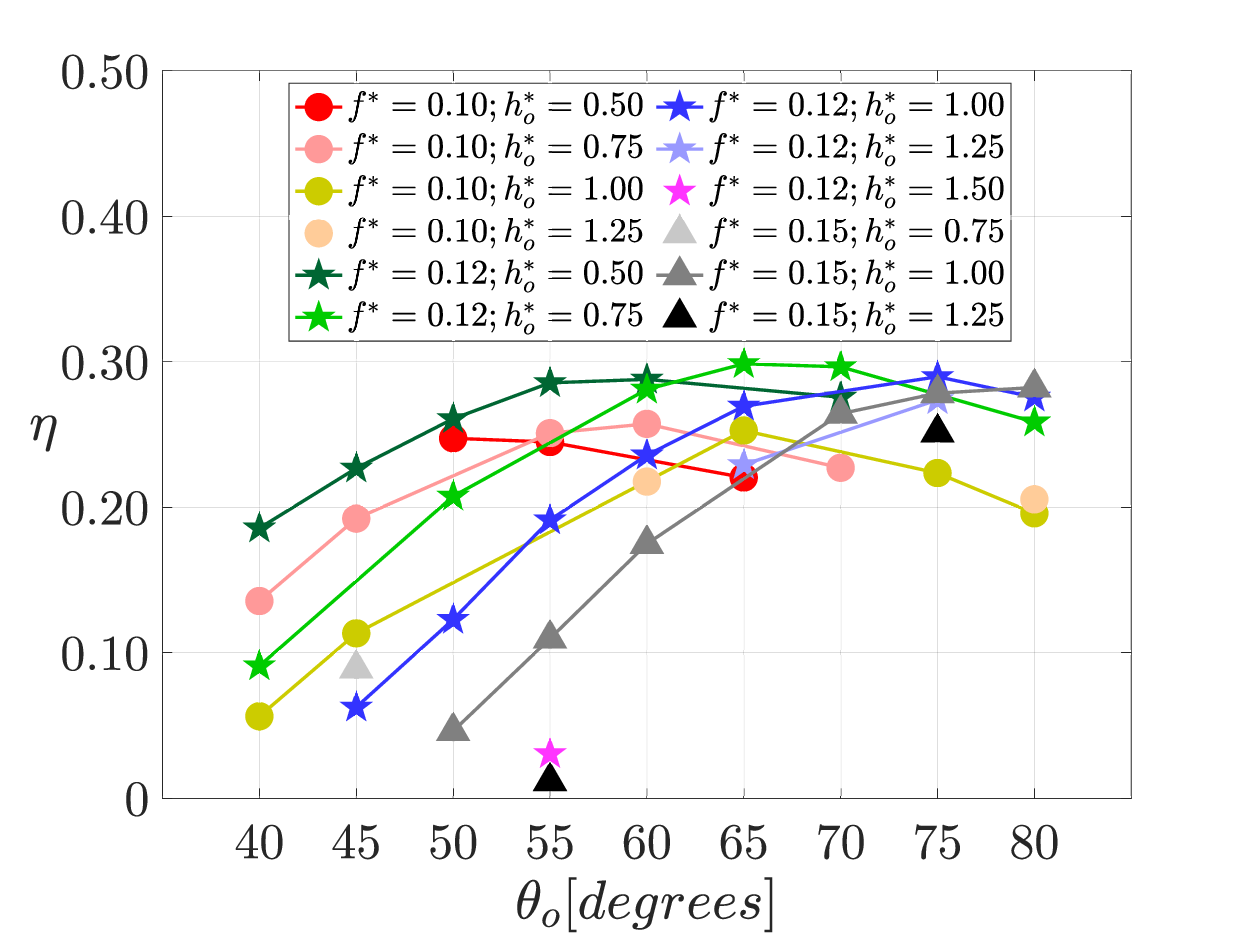}
        \caption{}
        \label{f:etaheave&pitch}
	\end{subfigure}
	\begin{subfigure}[b]{0.49\textwidth}
	\centering
        \includegraphics[width=1\textwidth]{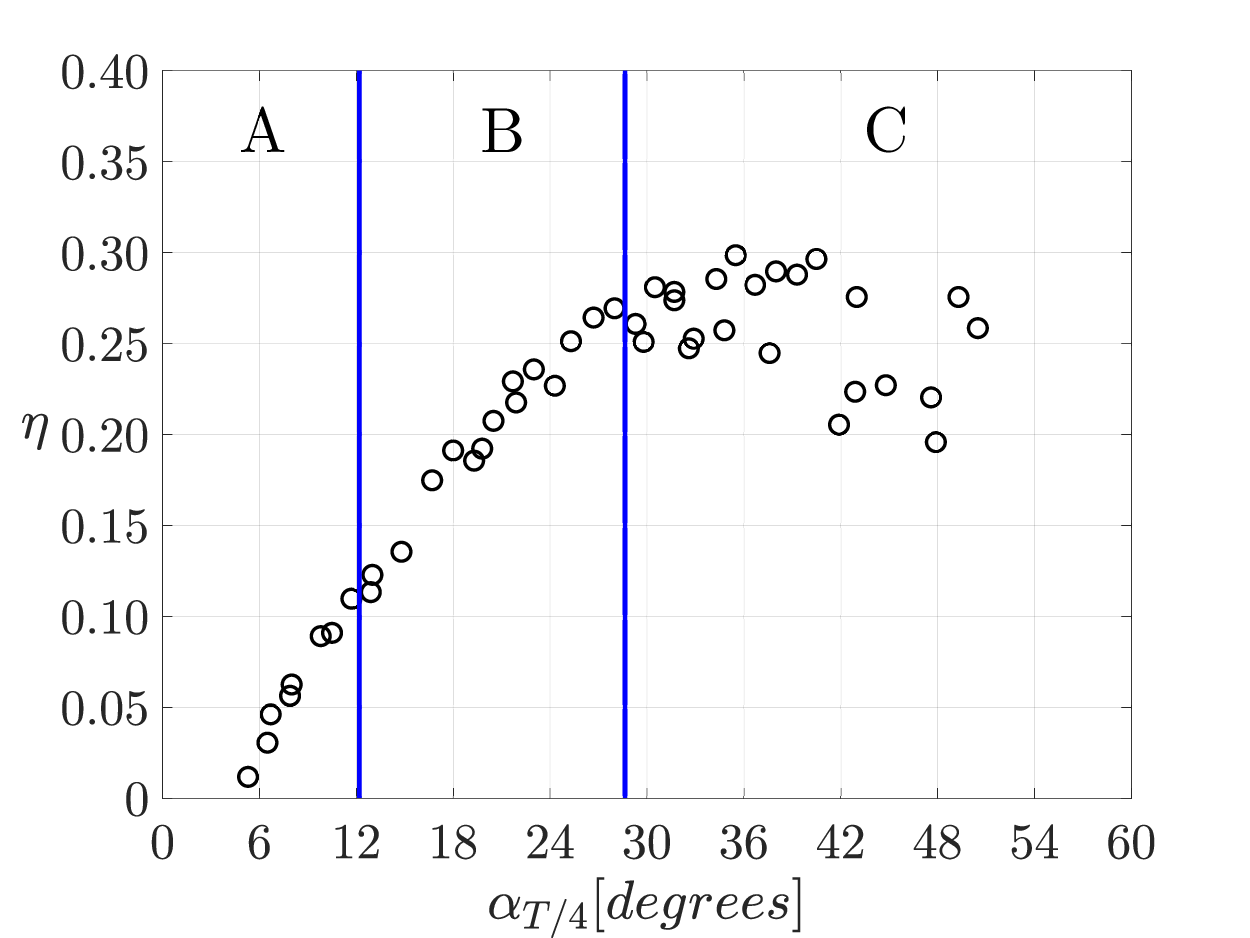}
        \caption{}
        \label{f:etaalpha}
	\end{subfigure}
\caption{(a) Energy harvesting efficiency, \bm{$\eta$}, as a function of pitch amplitude, \bm{$\theta_o$} (in degrees) for various reduced frequency (\bm{$f^*$}) and heave amplitude (\bm{$h_o^*$}) pairs; (b) Efficiency with respect to \bm{$\alpha_{T/4}$}. Vertical lines indicate initial grouping of vortex wake structures based on primary vortex strength analysis.}
\label{f:etaalpha_and_kins}	
\end{figure}

The kinematics outlined in Table \ref{t:kin} cover a large parameter space, which contributes to a range of energy harvesting modes, which range in efficiency from close to zero up to $30\%$. These results demonstrate that the optimal performance is not strongly correlated with a single set of foil kinematics. As displayed in Figure \ref{f:etaheave&pitch} high energy harvesting efficiency is found within the range $f^*=0.12-0.15; \theta_o=65^{\circ}-80^{\circ};h_o^*=0.50-1.00$ with no clear correlation with a single kinematic parameter. Thus, it is convenient to reduce the parameter space into a single and representative variable, $\alpha_{T/4}$, the characteristic relative angle of attack as defined in Equation \ref{eq:at4}, which allows foil efficiency to be expressed as a simpler function of foil kinematics as shown in Figure \ref{f:etaalpha} \cite{kindum2008}. The efficiency increases monotonically until $\alpha_{T/4}\approx28.0^{\circ}$ and then varies for higher $\alpha_{T/4}$ values due to a high degree of flow separation \cite{RibeiroFranck2020, RibeiroFranck2021}.

\begin{figure}[H]
\centering
\includegraphics[width=1\textwidth]{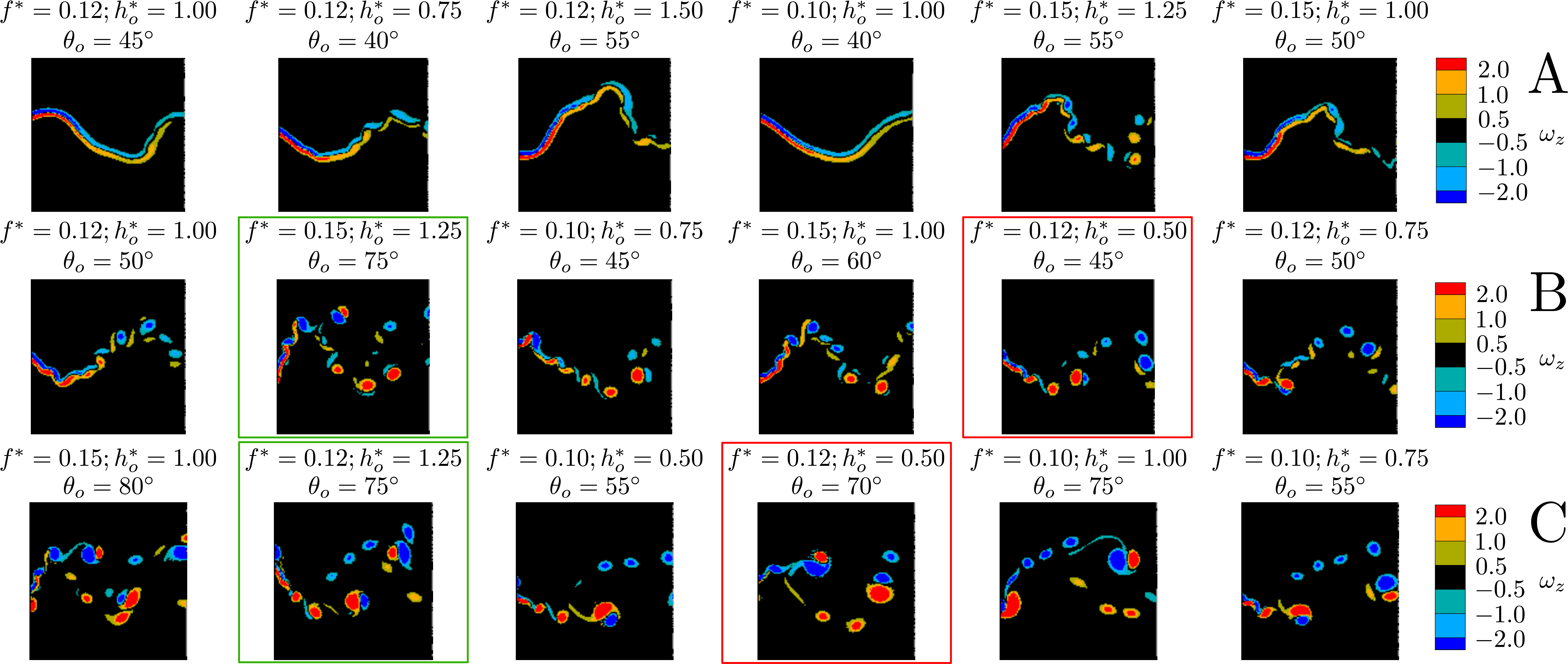}
\caption{Randomly selected images of wake structures from each class highlighted in Figure \ref{f:etaalpha}. The colored boxes outline wake features found when one foil parameter is varied while others remain constant. }
\label{f:visualwake}
\end{figure}

To correlate $\alpha_{T/4}$ with the flow structures that emanate from the foil, Ribeiro et al. \cite{RibeiroFranck2021} defined three classes labeled A ($5.3^{\circ} \leq\alpha_{T/4}\leq11.7^{\circ}$), B ($11.7^{\circ}<\alpha_{T/4}<29.3^{\circ}$) and C ($29.3^{\circ}\leq\alpha_{T/4}\leq50.5^{\circ}$) using the maximum strength of the primary vortex formed for each kinematics (as demonstrated in Figure \ref{f:etaalpha}). The wake structures that emerge from the three classes are displayed in Figure \ref{f:visualwake} within a $7.5c$ by $7.5c$ window located downstream, and general trends based on vortex strength are extracted from these wakes. Class A contains weak vortices as represented by a shear layer in the majority of images in this class. In contrast, class C displays the strongest and most coherent vortices compared with the other classes. Class B has a mix of kinematics with stronger vortices than class A but weaker than class C.

Observing the wake structures from Figure \ref{f:visualwake}, trends are found within and between classes. With an increase in pitch amplitude, vortices shed from the foil increase in size and strength with frequency and heave amplitude $f^*=0.12; h_o^*=0.50$ (see kinematics highlighted in red). With an increase in reduced frequency, the wake wavelength is smaller as displayed by the kinematics with $h_o^*=1.25, \theta_o=75^{\circ}$ (see kinematics in green).

\section{Supervised Classification} \label{s:classification}

In this Section, the wake is analyzed through an image-based supervised learning algorithm, where classes are defined based on the predetermined groupings motivated in Section \ref{s:vortexstrength}.

\subsection{Data Pre-Processing}

\begin{figure}[H]
\centering
\includegraphics[width=1\textwidth]{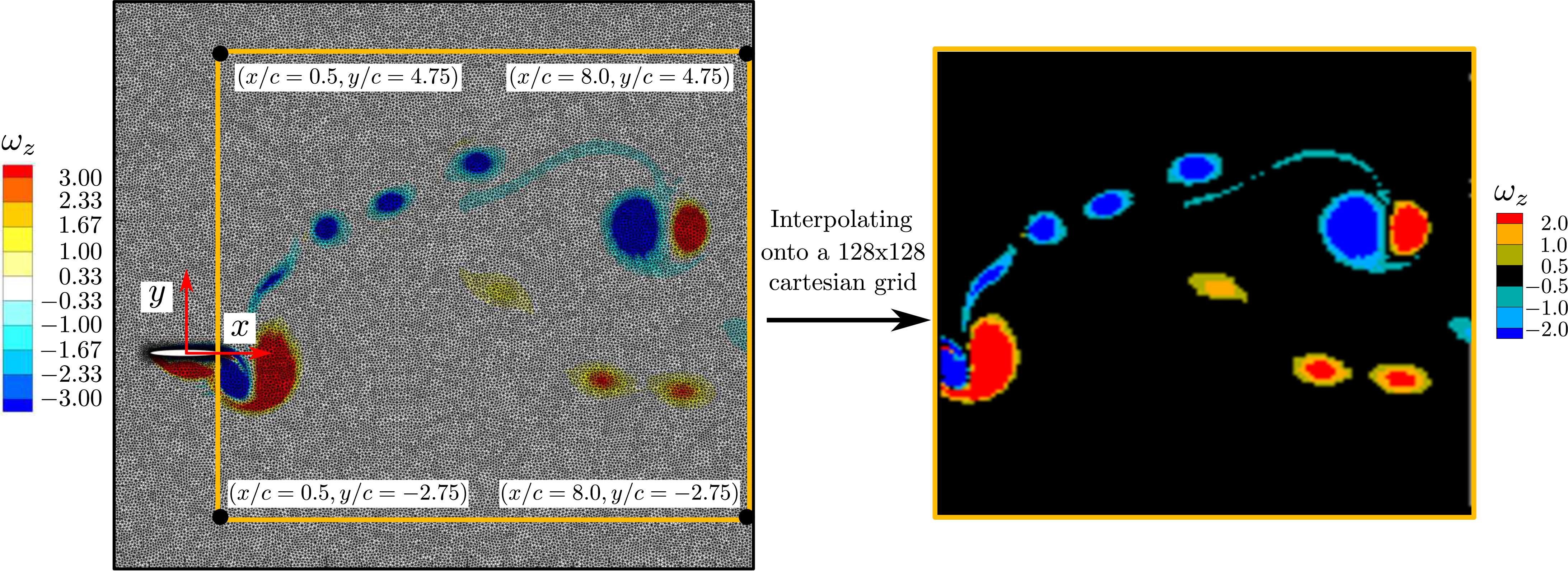}
\caption{Data pre-processing for the classification model interpolates the spanwise vorticity computed with high resolution computational data onto a 128x128 grid for a fixed region behind the foil at every 0.1 convective time units. }
\label{f:classification_preproc}
\end{figure}

The input to the classification model are images of 2D spanwise vorticity extracted from a $7.5c$ by $7.5c$ window located in the wake and interpolated onto a cartesian grid of $128$ by $128$ pixels as illustrated in Figure \ref{f:classification_preproc}. The window size is selected such that it contains all vortices shed from the foil in the $y$-direction in all kinematics from Table \ref{t:kin} and $7.5c$ corresponds to a typical inter-foil spacing in foil-arrays under the energy harvesting regime. For each set of kinematics, three oscillation cycles within the steady state regime are used as input with data sampled at every $tU_\infty/c=0.1$, for a total of $11,846$ samples. The fixed sampling rate is used in order to maintain the difference between consecutive wake images independent of the foil's reduced frequency, thus the number of samples differs for each frequency. Contour levels of vorticity are chosen to display the wake structures in all kinematics, and six levels ($-2,-1,-0.5,0.5,1,2$) are consistently drawn for each wake image as shown in Figure \ref{f:classification_preproc}.

Since the vortices shed from the foil may affect each other's trajectory \cite{RibeiroFranck2021, kindum2012, Ashraf2011}, the time evolution is considered in the classification neural network through the use of LSTM units. A sequence of five images is given as the input data, which provided higher accuracy compared with a sequence of ten images. To avoid overfitting, a data augmentation technique is also implemented, which duplicated the number of samples from $11,846$ to $23,692$. This technique not only took an input sequence at consecutive $0.1 tU_\infty/c$ units, but also with $0.1  tU_\infty/c$ skipped between samples, i.e. $0.1, 0.3, 0.5, 0.7, 0.9  tU_\infty/c$, following a similar strategy by Chong and Tay \cite{chong2017}.

\subsection{Classification Model Architecture} \label{s:modelarch}

\begin{figure}[htbp]
\centering
\includegraphics[width=1\textwidth]{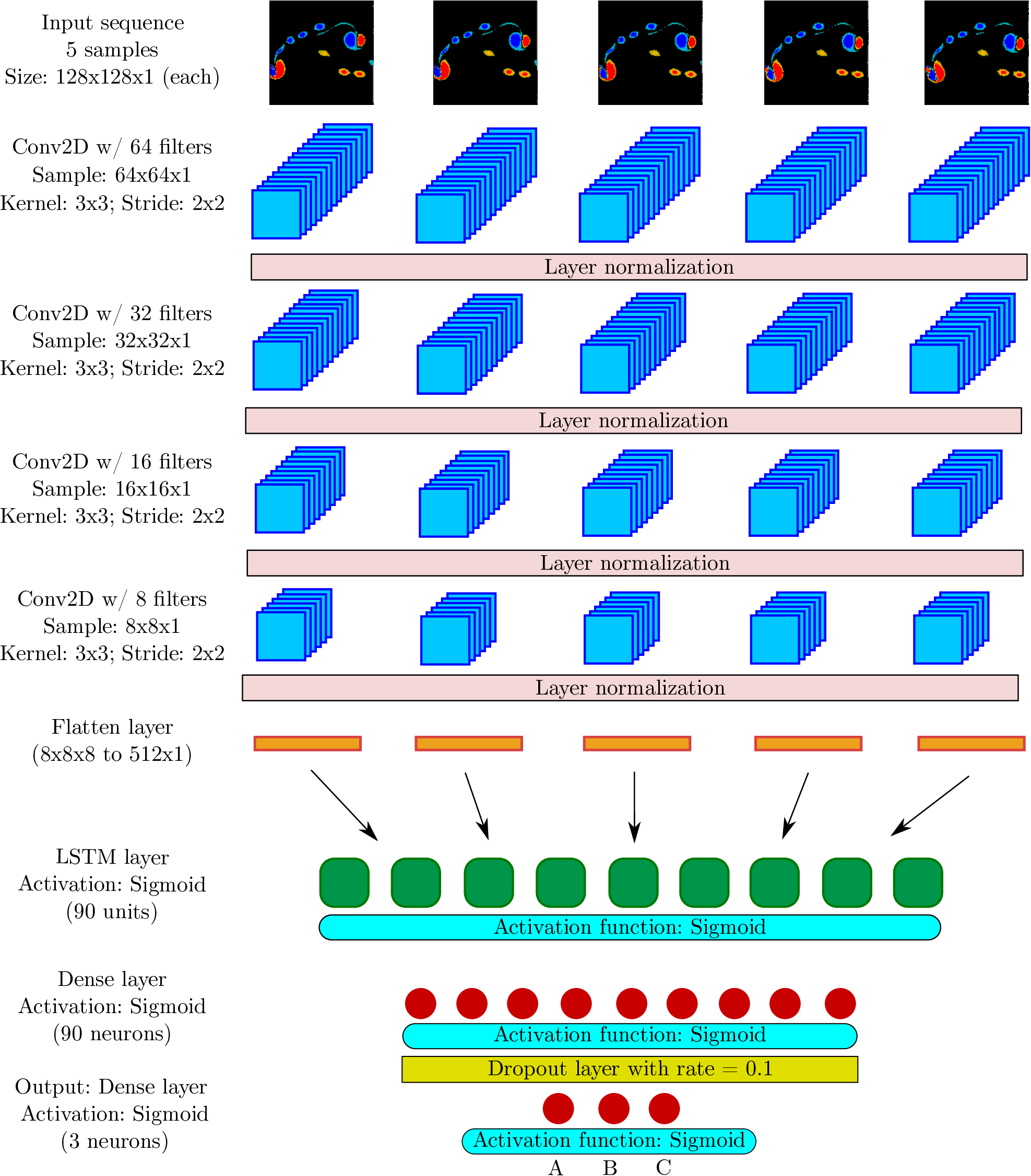}
\caption{Classification model architecture.}
\label{f:model}
\end{figure}

Using the Python based libraries TensorFlow \cite{tensorflow} and Keras \cite{keras}, the classification model is built from a combination of convolutional layers, LSTM units and dense layers, as outlined in Figure \ref{f:model}. The 2D convolutional layers (\textit{Conv2D}) are applied on each sample to extract the most significant features of the wake. These key features are detected with filters, which create feature maps through a convolutional operation on the preceding layer \cite{calvet2020}. Each convolutional layer uses a linear activation function with multiple filters of fixed kernel size, $3\times3$, that reduces the matrix dimensions while simultaneously keeping the most pertinent features. The number of feature maps define the depth of each convolutional layer and within each layer, a downsample operation is performed with a $2 \times 2$ stride, resulting in a reduction factor of $2$ in each matrix dimension while the depth remains constant. Each input sequence passes through four convolutional layers, decreasing the dimension of each sample from $128 \times 128 \times 1$ to an $8 \times 8$ feature map with eight channels.

Each sample is then flattened into a 1D vector with $512$ elements where $90$ LSTM units analyze the correlation between each wake image within the input sequence, with the goal of detecting patterns between each wake structure and its time evolution behind each foil configuration. The final section of the model contains a dropout layer of rate equals to $0.1$ that is placed between two dense layers in order to decrease overfitting. The dense layers classify each image according to the predetermined classes. The first layer contains $90$ neurons and the second dense layer has three neurons corresponding to class A, B or C. Both dense layers use a sigmoid activation function to normalize the output from the previous layer into a $0-1$ range. To update the neural network weights, the Adam optimization algorithm \cite{adam} is implemented in the model. To prevent overfitting, the early stopping technique \cite{prechelt1998} with $100$ training epochs is utilized.

The following hyperparameters in the classification model are tuned: number of filters in the convolutional layers, LSTM units, and number of neurons in the first dense layer. The sequences of $(64,32,16,8)$, $(32,16,8,4)$, $(128,64,32,16)$ filters for the convolutional layers are tested and $(64,32,16,8)$ obtained best performance. The LSTM units and number of neurons in the first dense layer are tuned using a range from $20-100$ units and neurons and it is found that $90$ units and $90$ neurons provided a higher model accuracy.

\subsection{Model Accuracy in Predicting Class Label of Test Data}

A five-fold stratified cross-validation is performed in the data, comprised of matrices of the vorticity fields and their respective class labels. This procedure helps decrease overfitting and it assures representative information of all prelabeled classes in each test subset \cite{fukami2020, bruntonbook, keras}.

Using the three classes defined in Section \ref{s:vortexstrength} (see Figure \ref{f:etaalpha}), the model is trained and an average accuracy of $92\%$ is obtained across the five folds. This indicates that approximately $92\%$ of all samples processed by the algorithm are labelled the same as their prelabels. 

\begin{figure}[H]
\centering
	\begin{subfigure}[b]{0.39\textwidth}
	\centering
        \includegraphics[width=0.8\linewidth]{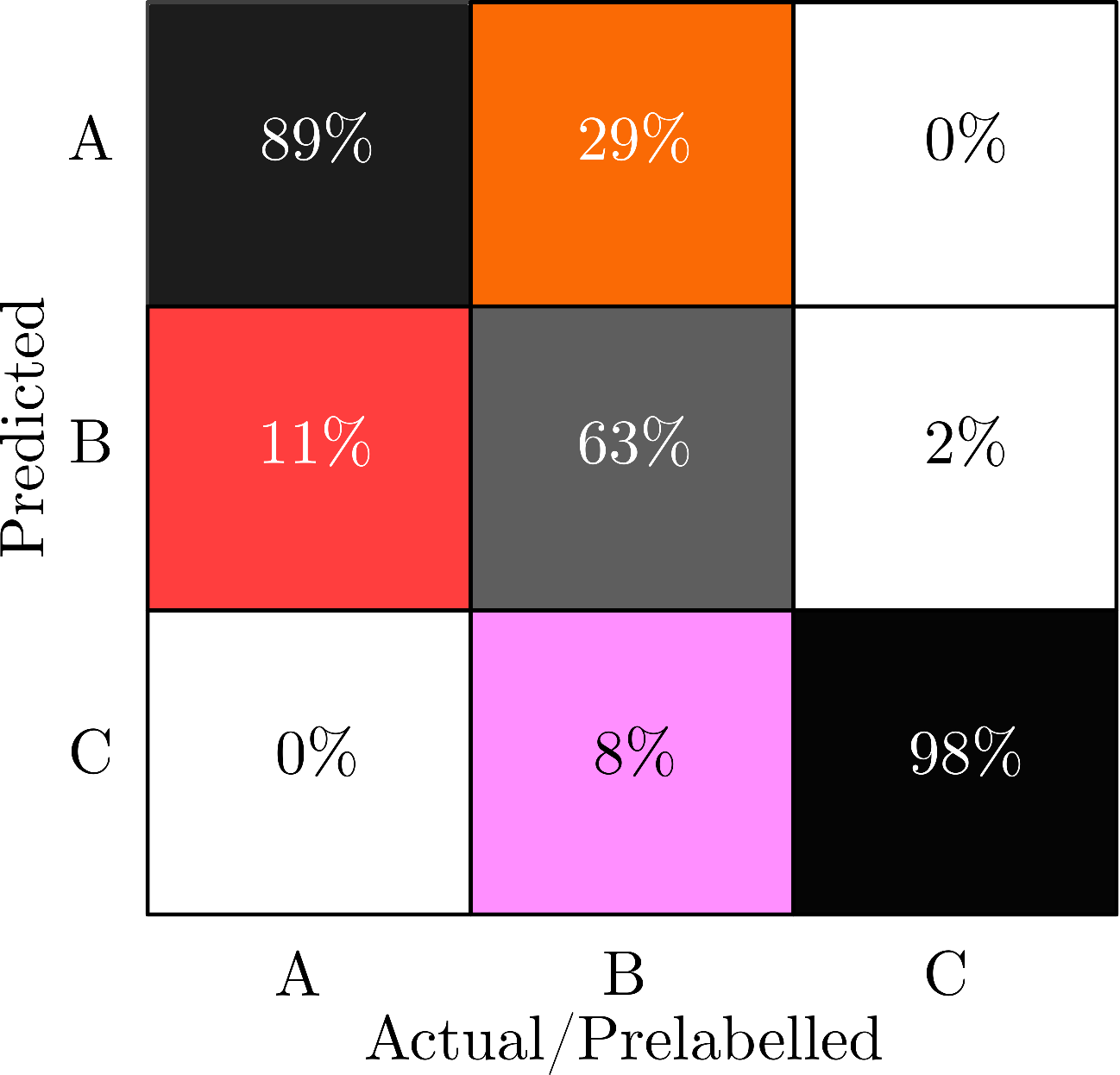}
        \caption{}
        \label{f:worst}
	\end{subfigure}
	\begin{subfigure}[b]{0.60\textwidth}
	\centering
        \includegraphics[width=0.9\textwidth]{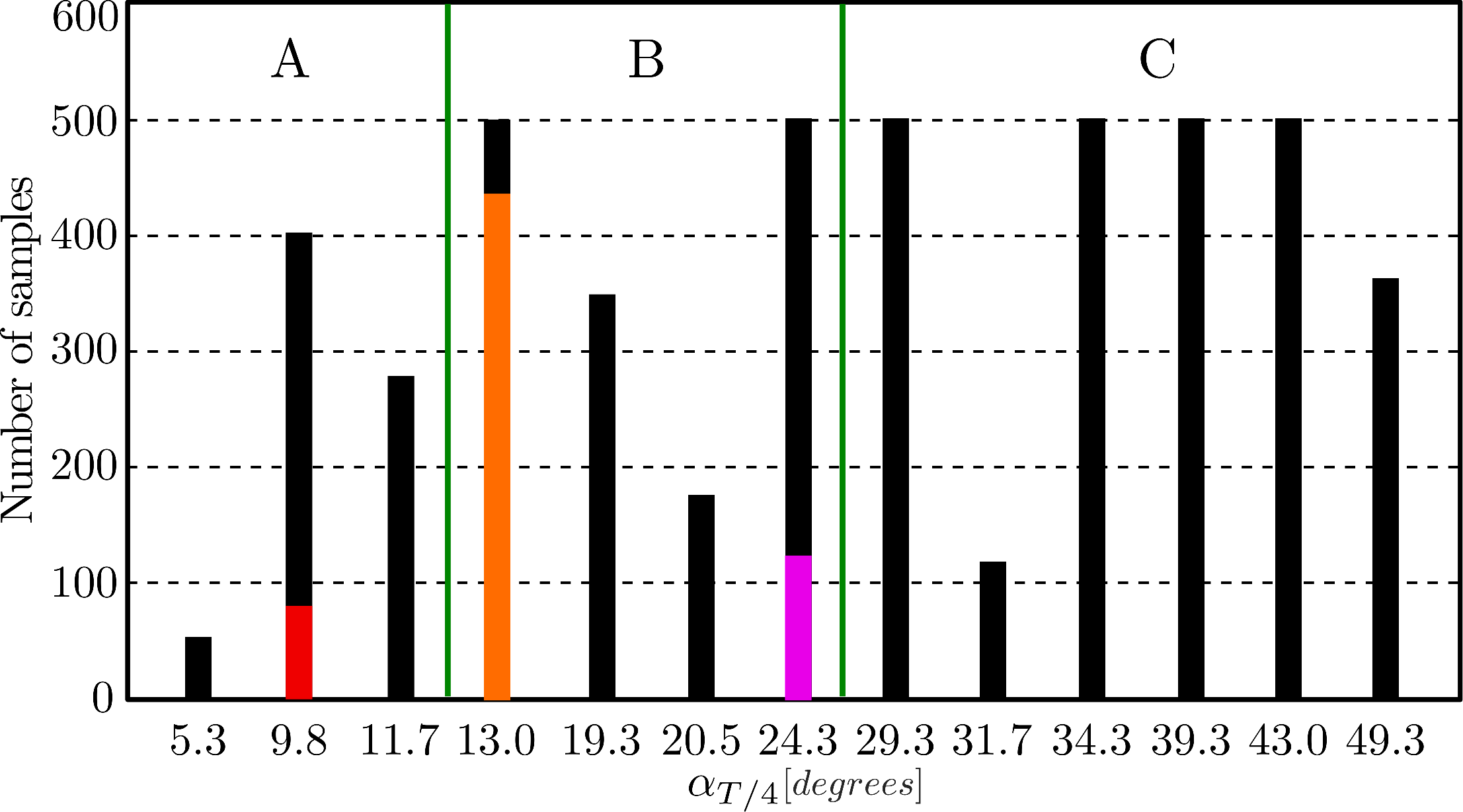}
        \caption{}
        \label{f:worstdist}
	\end{subfigure}
\caption{(a) Confusion matrix for the worst performance (fold \bm{$4$}) with mislabeled class A samples colored in red; mislabeled class B in orange and in pink. (b) Test set distribution of all foil kinematics in number of samples among \bm{$\alpha_{T/4}$} values presented in fold \bm{$4$}. Only a single foil kinematics had more than \bm{$50$\%} of images with a label mismatch between actual and predicted. The green lines divide the foil kinematics presented in each class.}
\label{f:worstanalysis3}	
\end{figure}

A confusion matrix, which is a common tool for summarizing the performance of a classification algorithm, and the corresponding mislabeled test data distribution for the worst performance fold (fold $4$) are shown in Figure \ref{f:worstanalysis3}. The algorithm does well discerning the kinematics within the class C but higher discrepancies are found for the classes A and B, with $11\%$ mislabeled samples in class A and $37\%$ in class B ($29\% + 8\%$). While the stratified cross-validation ensures a commensurate number of samples of each class in the training and test subsets, it does not ensure that each individual kinematics is represented in every test subset. The test data distribution in Figure \ref{f:worstdist} highlights the kinematics that have a label mismatch. The orange bar corresponds to the $29\%$ mislabeled samples that are classified as class A rather than class B, whereas red and pink bars show those mislabeled in classes B and C, respectively. Even though the mismatch percentages are high, there is only a single kinematics in fold $4$ ($\alpha_{T/4}=13.0^{\circ}$) that contains more than $50\%$ label mismatch, and the corresponding $\alpha_{T/4}$ is close to the class boundary between classes A and B ($\alpha_{T/4}=11.7^{\circ}$), which can explain the high mismatch.

\section{Unsupervised Clustering} \label{s:clustering}

The classification model presented in Section \ref{s:classification} divides the wake kinematics into three classes predetermined by the researcher (based on prior vortex analysis). In this Section, an unsupervised algorithm is utilized to group similar wake kinematics, and the clusters are subsequently compared with the classes from the supervised model. The unsupervised clustering groups the vortex images, individually, without any prior knowledge or relationship between wake kinematics and the respective wake structures. Therefore, the clusters obtained through this method will assist in verifying the classification results and potentially modifying the class boundaries determined by the researcher.

\subsection{Clustering Model Architecture and Convergence}

The unsupervised model architecture follows closely from the CAE clustering algorithm by Calvet et al. \cite{calvet2020}, including the same number of convolutional layers, filter and skip connections. The architecture consists of an autoencoder with five sequences of convolutional and max-pooling layers for the encoder portion. For the decoder portion, five sequences of convolutional and up sampling layers are used. The only hyperparameter retuned is the batch size, in which an online learning method (batch size equals to 1), is implemented within the autoencoder. With this algorithm there is no prelabeling of images, but the user must determine the number of clusters, which is explained below.

The input data is the same vorticity images described in Section \ref{s:classification}, except that there are no time sequences provided, and therefore each instantaneous wake image is treated independently. This results in $11,846$ unique samples from $46$ simulations. The training and validation data for the autoencoder utilize $27$ of the $46$ simulated kinematics. Within the $27$ kinematics, approximately $30\%$ of the samples are used for validation, corresponding to the first out of the three oscillation cycles, as outlined in Table \ref{t:kin}. Using samples of the same foil kinematics in the training/validation sets improved performance of the autoencoder while retuning. The remaining sets of foil kinematics ($19$) are used as test data after the autoencoder is tuned.

\begin{figure}[H]
\centering
\includegraphics[width=0.5\textwidth]{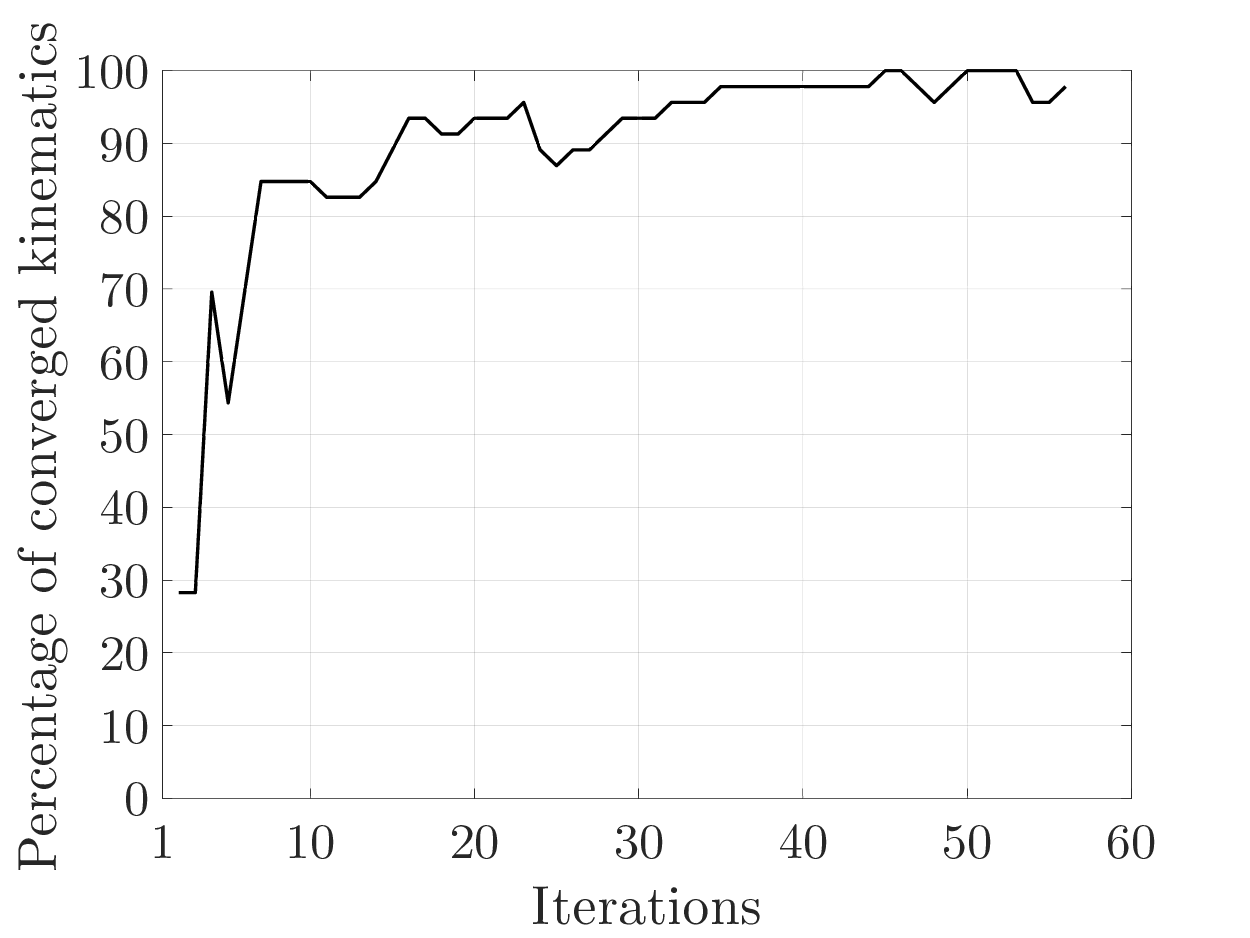}
\caption{Convergence of the autoencoder and clustering model.}
\label{f:convergence}
\end{figure}

Due to the stochastic nature of the model, the CAE clustering algorithm is trained for multiple iterations to check for convergence under a developed criteria. After each independent iteration, a set of kinematics is designated to the cluster that contains the majority of its samples (since samples are treated independently, samples from the same kinematics can end up in different clusters). For each iteration, the cluster label at which a set of kinematics is assigned to is recorded. Convergence is reached if a set of kinematics remains in the same cluster (same label) as the previous iteration. Figure \ref{f:convergence} shows the percentage of foil kinematics that converged when four clusters are considered. A total of $56$ random clusterings is performed and after $35$ iterations, the algorithm consistently maintains convergence of at least $96\%$ of all foil kinematics being assigned to a unique cluster.

\begin{figure}[H]
\centering
        \includegraphics[width=0.5\linewidth]{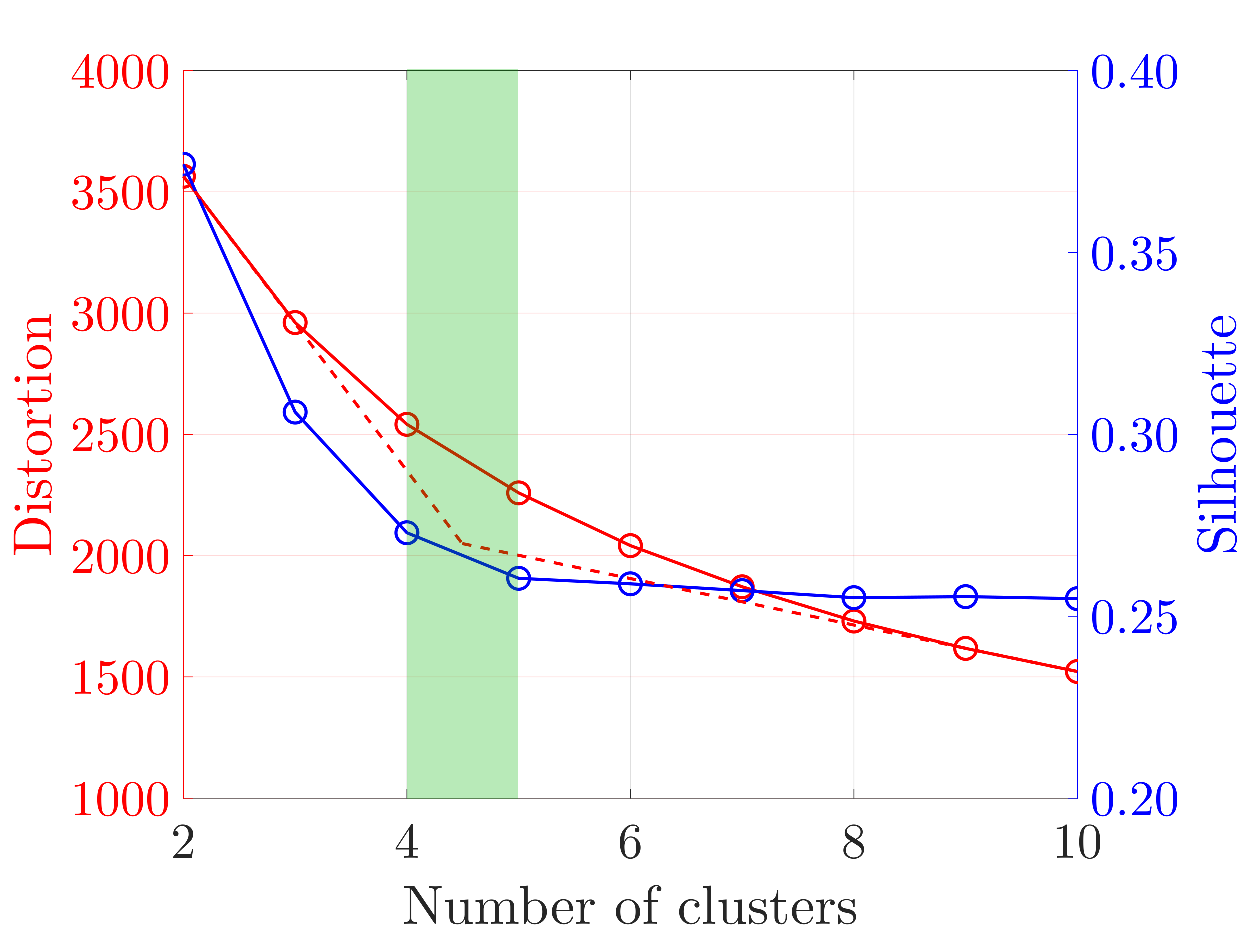}
\caption{Distortion (red solid line) and silhouette score (blue solid line) with respect to number of clusters and averaged over all iterations. The red dashed lines are tangent to both ends of the distortion line and their intersection point corresponds to the distortion curve's elbow. The green shaded region displays the numbers of clusters that are closest to the elbow.}
\label{f:elbowsilhouette}	
\end{figure}

A combination of the elbow and the silhouette score methods are utilized to determine the optimal number clusters for the provided samples, following the approach from Calvet et al. \cite{calvet2020}. The elbow method \cite{elbow} computes the total within-cluster sum of square error, known as distortion, and the number closer to the ‘elbow’ of the curve is the indicator of the approximate number of clusters that best separates the data. The silhouette score method \cite{silhouettes} determines how well each image lies within its cluster by estimating cohesion (intra-class) and separation (inter-class) as Euclidean distances. The score is a combination of both factors, and ranges from zero to one, where a higher score indicates better clustering. The distortion and silhouette scores are averaged over all iterations with results displayed in Figure \ref{f:elbowsilhouette}. The elbow, represented by the intersection point of the tangent dashed lines to both ends of the distortion curve in Fig. \ref{f:elbowsilhouette}, is located between four and five clusters (see green shaded region) but a slightly higher silhouette score is found when the data is divided into four clusters. Therefore, the optimal number of clusters is set to four. Due to the small differences in the averaged silhouette score, an analysis is also performed using five clusters but does not provide any additional physical insights on wake patterns compared with four clusters.

\subsection{Updating Class Boundaries Using the Clustering Results}

\begin{figure}[H]
\centering
	\begin{subfigure}[b]{0.49\textwidth}
	\centering
        \includegraphics[width=1\linewidth]{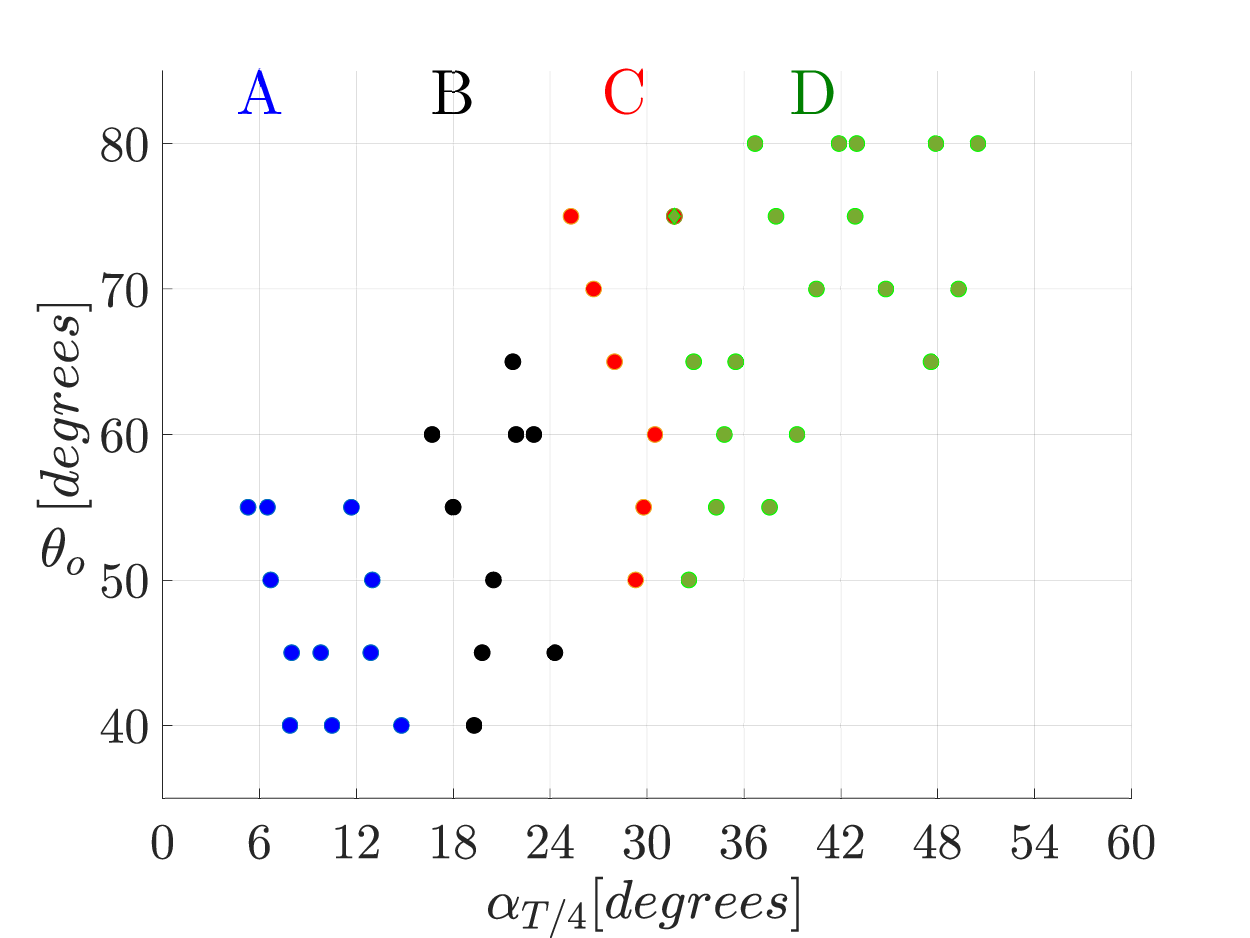}
        \caption{}
        \label{f:clustergroups}
	\end{subfigure}
	\begin{subfigure}[b]{0.49\textwidth}
	\centering
        \includegraphics[width=1\textwidth]{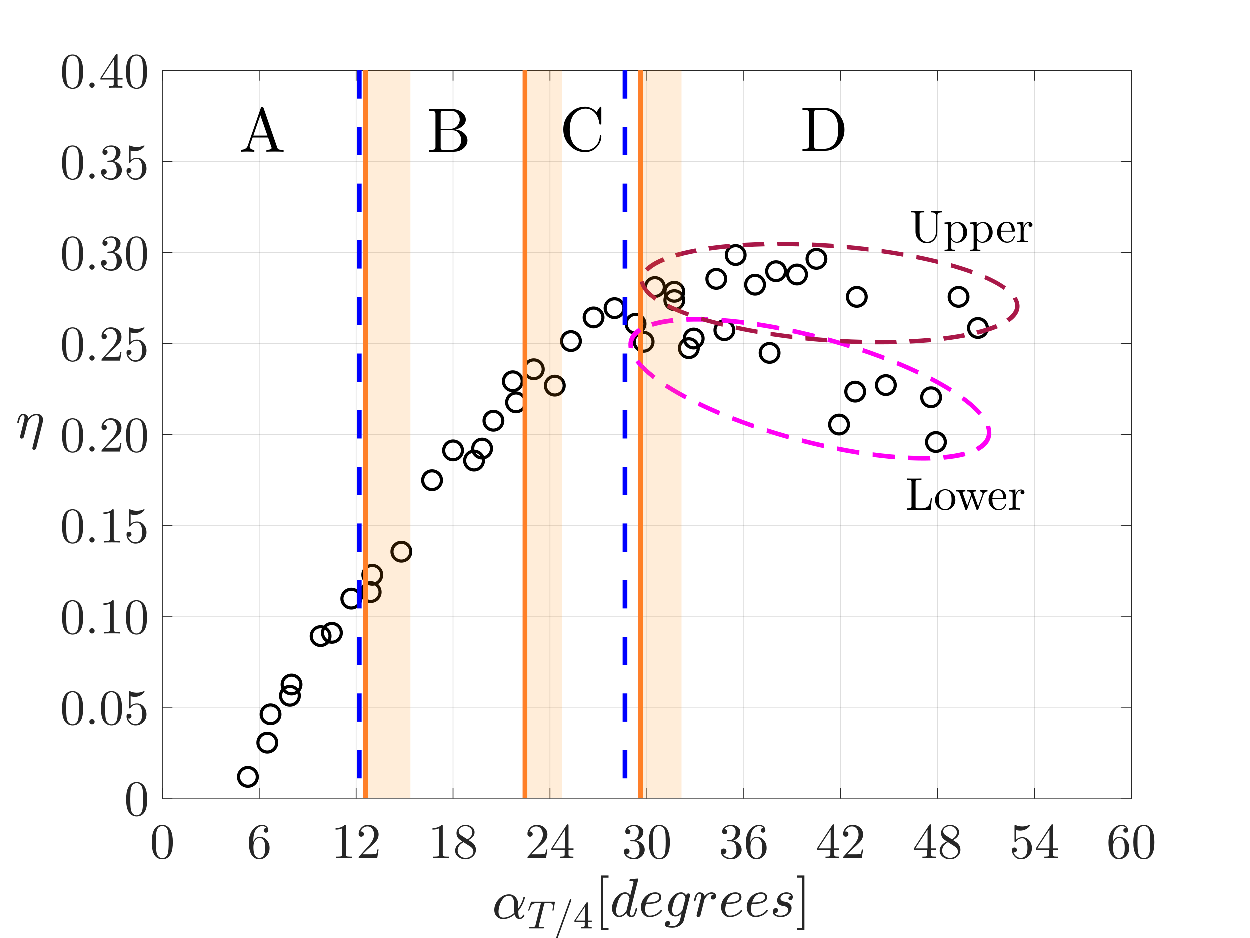}
        \caption{}
        \label{f:updatedgroupseta}
	\end{subfigure}
\caption{(a) Clustering results represented by each foil kinematics as a data point: two kinematics with \bm{$\alpha_{T/4}=31.7^{\circ}$} and \bm{$\theta_o=75^{\circ}$} represented by a red circle and a green diamond placed in clusters C and D, respectively; (b) Updated class divisions (orange solid lines) compared with original groupings (blue dashed lines) along with the mismatch between class and cluster boundaries (orange shaded regions).}
\label{f:updatedboundaries}	
\end{figure}

The results of four clusters are displayed in Figure \ref{f:clustergroups}. Every cluster is roughly defined within an $\alpha_{T/4}$ range, which is consistent with the results from the supervised model that also utilized $\alpha_{T/4}$ as the preferential kinematic parameter. The clustering division of four clusters, rather than three, naturally imposes a new boundary and introduces small shifts in the other two cluster boundaries. These new divisions are explored by utilizing the previously described classification algorithm in the following manner. 

For instance, the cluster A groups foil kinematics in the range of $\alpha_{T/4} \leq 14.8^{\circ}$ (Figure \ref{f:clustergroups}), whereas the previous class A included kinematics with $\alpha_{T/4}\leq 11.7^{\circ}$. The proposed clustering division is implemented within the classification algorithm previously described (with the five-fold cross-validation). The results were less accurate, indicating that the classification algorithm can more reliably group images when the boundary is closer to $\alpha_{T/4}\leq 11.7^{\circ}$.

Similar tests are performed on the remaining two cluster-informed divisions at $\alpha_{T/4}=24.3^{\circ}$ and $\alpha_{T/4}=31.7^{\circ}$. The results indicate that shifting both these boundaries to slightly lower $\alpha_{T/4}$ reveals more accurate classification of images. These results inform the update on the class boundaries as proposed by the orange solid lines in Figure \ref{f:updatedgroupseta}, which are contrasted with the original classifications from Section \ref{s:classification} (dashed blue lines). The orange shaded regions next to each boundary represent the mismatch between the class and cluster boundaries.

To summarize, when four classes are defined, the highest classification accuracy is obtained when the class boundaries are placed at $\alpha_{T/4}=11.7^{\circ}$, $\alpha_{T/4}=23.0^{\circ}$ and $\alpha_{T/4}=29.3^{\circ}$ (orange lines from Figure \ref{f:updatedgroupseta}). The results of these tests reveal that the accuracy is just as good as the original three classes, with an average fold accuracy of $91\%$. Furthermore, mismatched labels only occur close to the boundary divisions. For instance, $100\%$ of the samples from class A in Folds $1$ and $2$ have a label match between predicted and prelabelled, with the same occurring in Fold $4$ for classes B and C. For Fold $1$, only the foil kinematics with $\alpha_{T/4}=28.0^{\circ}$ is mislabelled between classes C and D, which can be explained by the proximity of this kinematics to its neighboring class. All other arrangements of class boundaries that are tested yield an average accuracy lower than $91\%$.

\begin{figure}[H]
\centering
	\begin{subfigure}[b]{0.39\textwidth}
	\centering
        \includegraphics[width=1\linewidth]{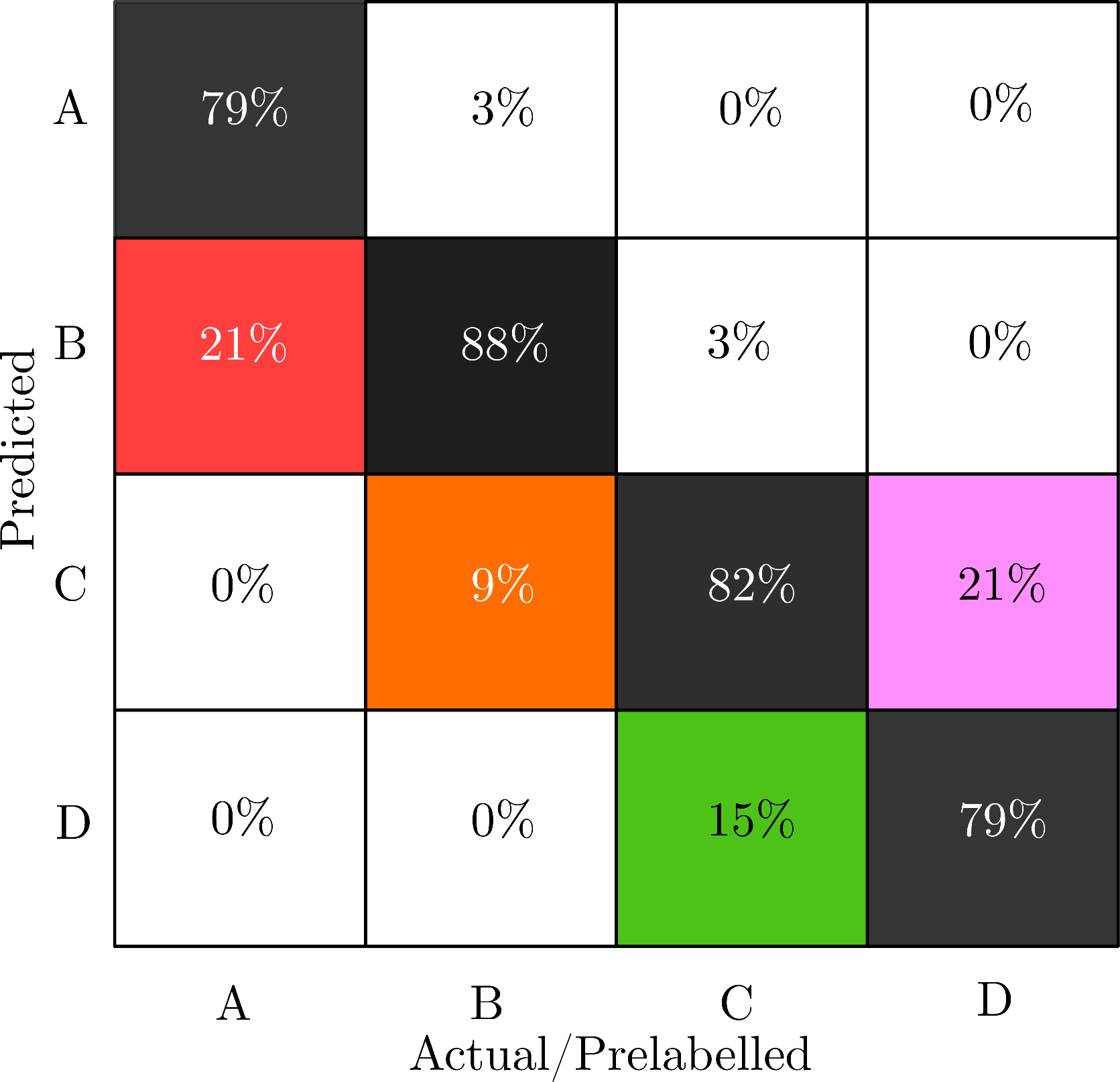}
        \caption{}
        \label{f:worst4}
	\end{subfigure}
	\begin{subfigure}[b]{0.60\textwidth}
	\centering
        \includegraphics[width=0.9\textwidth]{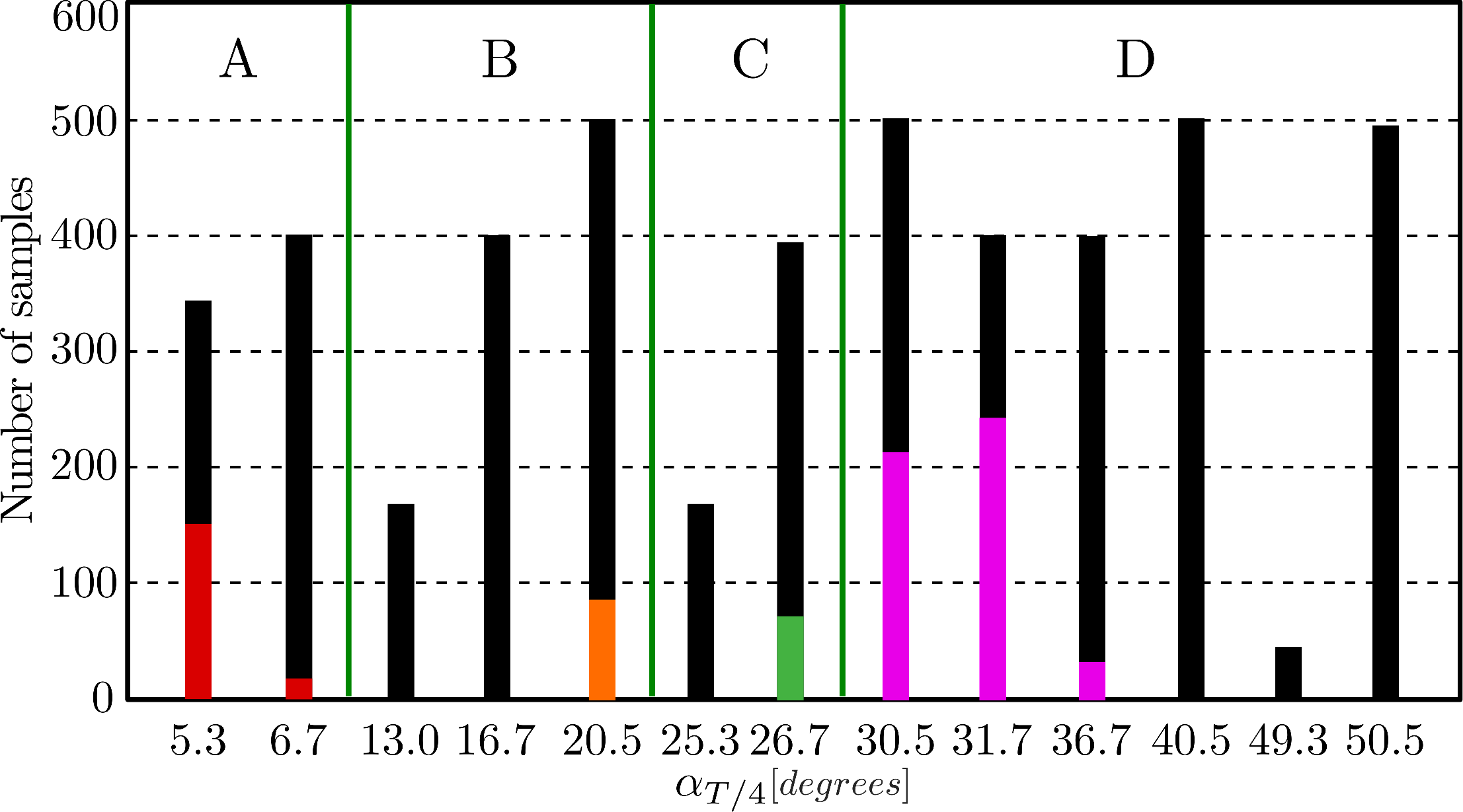}
        \caption{}
        \label{f:worstdist4}
	\end{subfigure}
\caption{(a) Confusion matrix for the worst performance (fold \bm{$5$}) with mislabeled class A samples colored in red; mislabeled class B in orange; mislabeled class C in green; mislabeled class D in pink. (b) Test set distribution of all foil kinematics in number of samples among \bm{$\alpha_{T/4}$} values presented in fold \bm{$5$}. The kinematics \bm{$f^*=0.15; h_o^*=1.00; \theta_o=75^{\circ}$} is the only foil kinematics that had more than \bm{$50$\%} of images with a label mismatch between actual and predicted. The green lines divide the data presented in each class.}
\label{f:worstanalysis4}	
\end{figure}

To illustrate the kinematics that are commonly mislabeled by the algorithm, the confusion matrix for the worst performance fold (fold $5$) is displayed in Figure \ref{f:worst4}. Even for this scenario, at least $79\%$ of samples are properly labelled. Those not accurately predicted are shown in Figure \ref{f:worstdist}, with only a single foil kinematics with $\alpha_{T/4}=31.7^{\circ}$ ($f^*=0.15; h_o^*=1.00; \theta_o=75^{\circ}$) containing a label mismatch higher than $50\%$. An outlier is the kinematics with $\alpha_{T/4}=5.3^{\circ}$ ($f^*=0.15; h_o^*=1.25; \theta_o=55^{\circ}$) which contains a label mismatch lower than $50\%$ and is not close to a neighbouring class. A possible explanation is the roll-up of the shear layer into multiple weak vortices generates a wake trajectory that is confused with the wake pattern of a higher $\alpha_{T/4}$. The wake structures found in this foil kinematics are displayed in Figure \ref{f:visualwake}.

\begin{figure}[H]
\centering
        \includegraphics[width=1\linewidth]{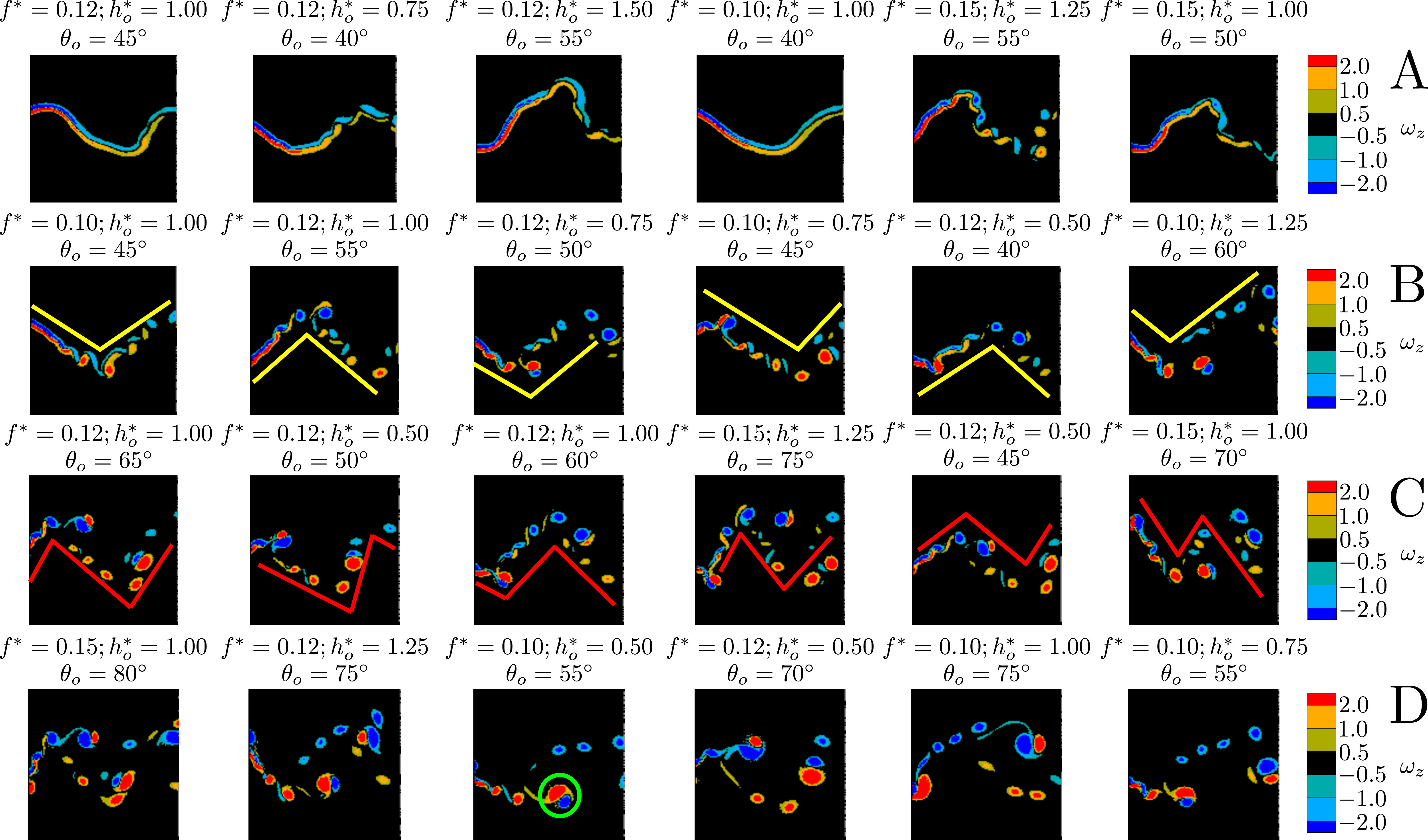}
\caption{Wake structures colored by spanwise vorticity (\bm{$\omega_{z}$}) for each class displayed in Figure \ref{f:updatedgroupseta}. The wake images are randomly selected within each class and the foil kinematics corresponding to each wake is displayed at top of each image. The yellow, red lines and green circle correspond to each wake pattern.}
\label{f:clusteringwakes}	
\end{figure}

The four updated classes offer new physical insight on wake patterns, illustrated in Figure \ref{f:clusteringwakes}. Each row highlights wake images at different foil positions that are randomly selected from various kinematics within each class. In class A ($5.3^{\circ} \leq \alpha_{T/4}\leq 11.7^{\circ}$), the foil generates a shear layer wake pattern as noticed by the absence of coherent vortices in the wake, and as previously described in Section \ref{s:vortexstrength}. Although classes B ($11.7^{\circ}<\alpha_{T/4}<23.0^{\circ}$) and C ($23.0^{\circ} \leq \alpha_{T/4} \leq29.3^{\circ}$) contain stronger vortices, the wake path is considerably different, with class B showing a longer wavelength within the selected wake window compared to class C (see yellow and red lines), which is a feature not previously captured by the original class divisions. Class D ($29.3^{\circ}<\alpha_{T/4}\leq50.5^{\circ}$) contains kinematics where the foil generates the largest number and strongest coherent vortices, as shown by the presence of a strong primary LEV in the majority of wake images (see green circle). These differences in the wavelength among wake patterns emphasize the criteria used in unsupervised clustering to differentiate each regime which are not captured by the initial classification approach.

The wake patterns obtained by the classification and clustering models are significantly different from those found in propulsive foils. Comparing with previous work by Calvet et al. \cite{calvet2020}, they obtained six different wake patterns that are correlated with two parameters, $\alpha_{T/4}$ and Strouhal number $\left(St=\frac{2fh_o}{U_\infty}\right)$, which varied from $0^{\circ}<\alpha_{T/4}<40^{\circ}$ and $0.2<St<1.2$. Due to the higher reduced frequency of propulsive foils, more wake structures are found closer to the foil and this contributes to a higher contrast of wake patterns across clusters compared with those in Figure \ref{f:clusteringwakes}. This contrast also contributes to the difference in the number of wake patterns between foil regimes. While the work by Calvet et al. obtained six wake patterns behind propulsive foils, the analysis performed here identified four distinct wakes under the energy harvesting regime.

Class D demonstrates high variation in efficiency within its kinematics. This is observed between $\alpha_{T/4}=40.5^{\circ}$ and $\alpha_{T/4}=41.9^{\circ}$ where efficiency drops by approximately $9\%$ (see Figure \ref{f:updatedgroupseta}). This could be described as a bifurcation in the efficiency curve at $\alpha_{T/4}>28.0^{\circ}$ as illustrated by an upper and lower branches. However, neither the classification nor clustering models could discern differences in the wakes between the higher and lower branches of efficiency within class D. To further investigate these branches, the foil parameters corresponding to each kinematics are explored and it is noticed that all foil kinematics in the lower branch have a reduced frequency of $fc/U_\infty=0.10$ and the upper branch, $fc/U_\infty=0.12-0.15$.

\begin{figure}[H]
\centering
        \includegraphics[width=0.5\linewidth]{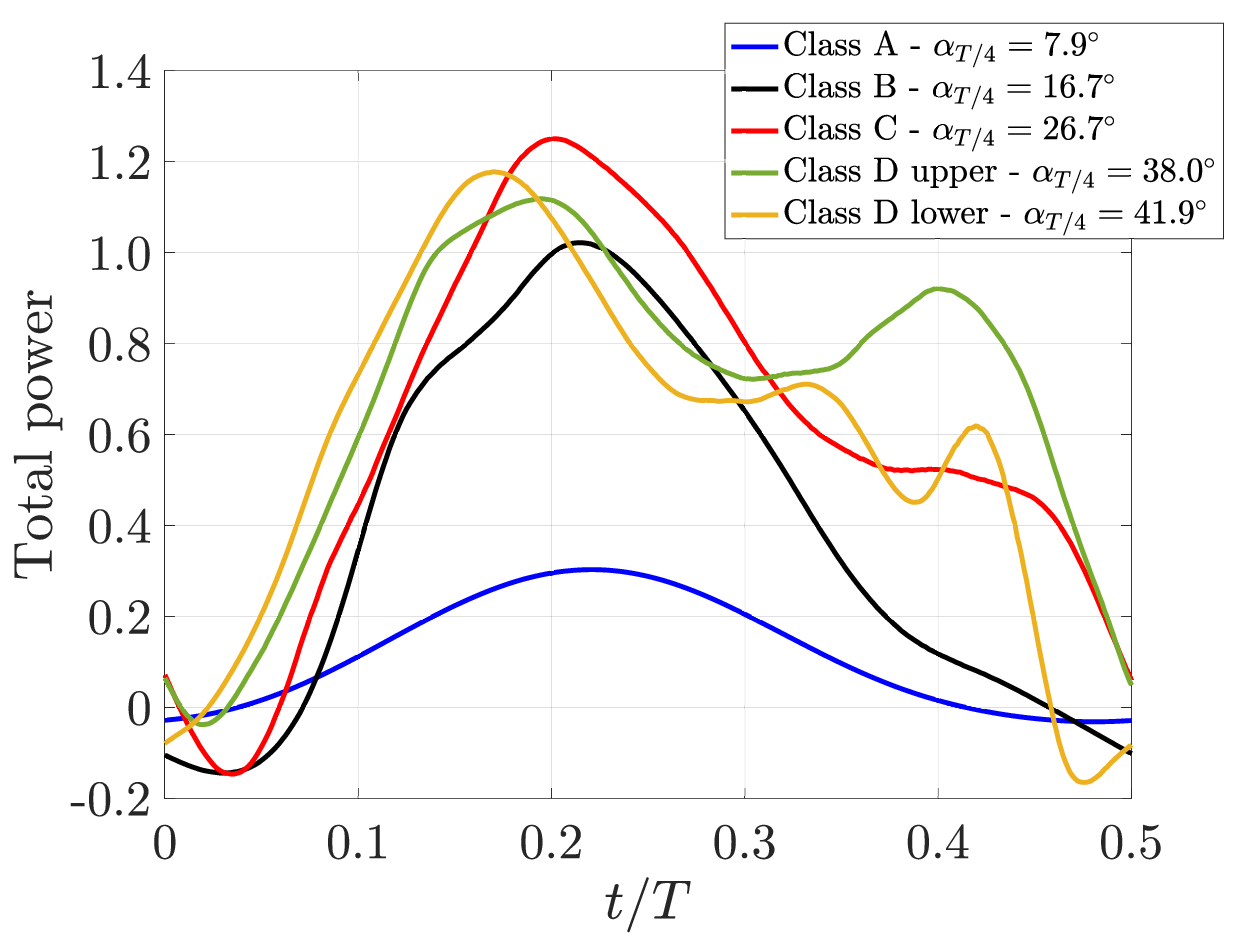}
\caption{Phase-averaged total power extracted in a half-cycle from a representative foil kinematics within each class.}
\label{f:powertrends}	
\end{figure}

The new updated classes also provide patterns in power extraction, as displayed in Figure \ref{f:powertrends}. Each curve corresponds to the phase-averaged total power extracted in a half-cycle from a representative foil kinematics in each class. All classes display a power peak close to the mid-stroke position ($t/T=0.25$), which corresponds to the foil's maximum heave velocity and thus typically is where maximum power is reached. Class A shows a smooth power profile with a lower amplitude compared to the other classes as expected due to the absence of coherent vortices generated by the foil. With the formation of LEVs as $\alpha_{T/4}$ increases, class B still highlights a smooth profile and class C indicates a higher power magnitude and higher unsteadiness on $t/T=0.3-0.5$. This unsteady behavior is most likely caused by secondary vortices formed on the foil due to a higher $\alpha_{T/4}$ in class C compared to class B. This unsteadiness is more apparent in the lower branch of class D where large and strong vortices are formed and shed from the foil. The power profile in the representative kinematics of the upper branch is similar to the lower branch in the region $t/T=0-0.3$ but it displays a second power peak in the remaining portion of the half-cycle. This peak is caused by the higher reduced frequency of the kinematics in the upper branch, which contributes to a delay in the vortex shedding and thus more power is extracted.

\begin{figure}[H]
\centering
        \includegraphics[width=1\linewidth]{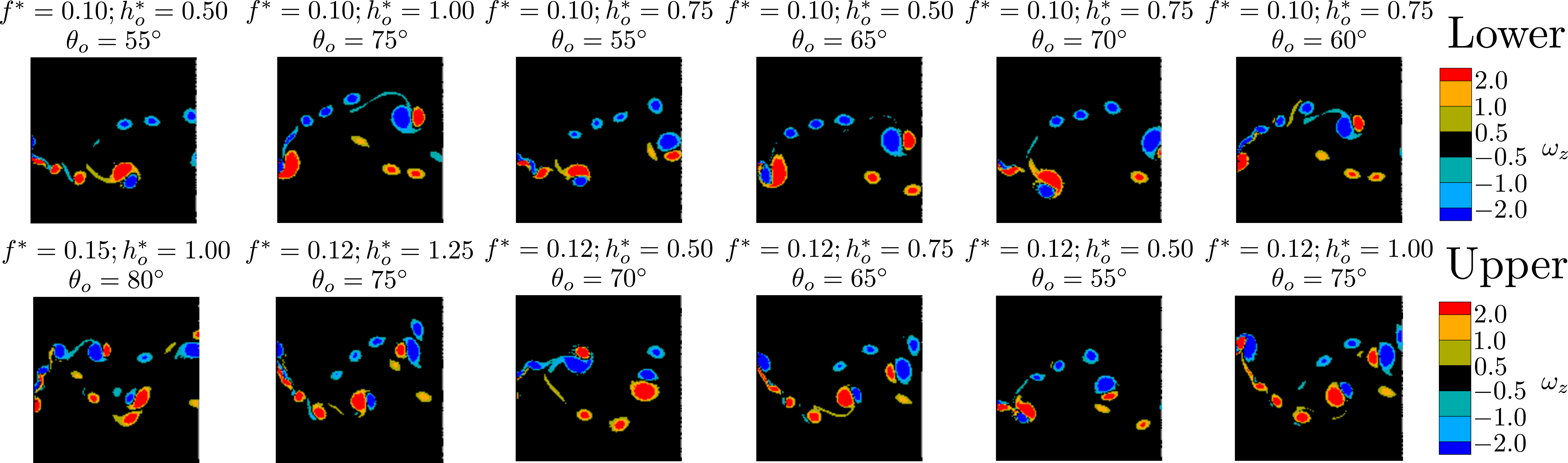}
\caption{Wake structures colored by spanwise vorticity (\bm{$\omega_{z}$}) that emerge from foil kinematics within each branch in class D highlighted in Figure \ref{f:updatedgroupseta}. The wake images are randomly selected and the foil kinematics corresponding to each wake is displayed at top of each image.}
\label{f:brancheswakes}
\end{figure}

To visualize those vortex structures, the wakes from both upper and lower branches are illustrated in Figure \ref{f:brancheswakes}. As observed in Section \ref{s:vortexstrength} and by Ribeiro et al. \cite{RibeiroFranck2020}, a lower reduced frequency is correlated with vortex structures spending less time on foil surface, which decreases the pressure gradient around the foil. The wake structures between upper and lower branches are different with more vortices located within the wake window in the upper branch due to higher foil's reduced frequency but still no pattern can be visualized. Although neither the classification nor clustering models could discern the differences just described in these branches, a possible solution would be to provide additional information about the kinematics of each wake image to the convolutional neural network like the reduced frequency, similar to the method implemented by Morimoto et al. \cite{morimoto2021}, but it is not explored in this investigation. Another potential solution would be to have more foil kinematics and hence more data in class D.

The wake patterns obtained from the updated classes can also be used to predict the power extraction of foil-arrays. For instance, the wake pattern generated from a class with low $\alpha_{T/4}$ values ($5.3^{\circ} \leq \alpha_{T/4}\leq 11.7^{\circ}$) does not significantly impact energy extraction from foils placed downstream due to the absence of coherent vortices that disturbs the oncoming flow of trailing foils. The opposite is true for a class with high $\alpha_{T/4}$ values ($29.3^{\circ}<\alpha_{T/4}\leq50.5^{\circ}$).

\section{Conclusion} \label{s:conclusion}

A machine learning model is developed to classify wake structures behind an oscillating foil in the energy harvesting regime of flapping foil kinematics. The goal of the paper is to utilize the machine learning algorithm to sort and classify wake modes using the vorticity fields downstream of the oscillating foil and correlate the kinematics with associated wake patterns. This model gives insight on wake similarity among various foil kinematics, which is important to build predictive models of oscillating foil arrays for energy harvesting.

Data is obtained through simulations of oscillating foils at $46$ unique kinematics, and time-dependent vorticity flow fields are extracted at equal times across three simulation cycles to form a total of $23,692$ samples. Based on previous work \cite{RibeiroFranck2021}, three initial classes are defined based on values of the relative angle of attack, $\alpha_{T/4}$. The classification model consists of four convolutional layers and $90$ LSTM units applied on multiple input sequences of five samples each. The model's output consist of three neurons corresponding to the classes A, B, C. After the model is trained and tuned, the average test accuracy among all folds is $92\%$ with the majority of foil kinematics showing a label mismatch percentage less than $50\%$ between actual and predicted, demonstrating the model's ability to discern wakes among classes.

Although the classification model is successful in finding wake patterns the class divisions are predetermined by the researcher, and assumed to correlate with only one parameter, $\alpha_{T/4}$, thus biasing the relationship between wake structure and flapping kinematics. Therefore, an unsupervised approach is performed through a CAE clustering algorithm, which does not require any prelabelling or bias. The results indicate that there is still a strong correlation with $\alpha_{T/4}$, and that clusters naturally align with this kinematic parameter. Furthermore, analysis shows that four clusters are optimal instead of three that were originally proposed. 

In summary, the clustering model provided validation that $\alpha_{T/4}$ was a predictive kinematic parameter for wake structure, and outlined an additional grouping previously undetected by the researcher. A final configuration of four new classes is proposed which results in an average fold accuracy of $91\%$ using the classification algorithm. The four classes offer new physical insight on wake patterns within each range of foil kinematics based on vortex strength and oscillatory wake structure. Further analysis is performed in the class with the highest $\alpha_{T/4}$ values and additional wake patterns are obtained that could not be captured by either the classification or clustering algorithms. This research builds upon the knowledge of how wake patterns and kinematics are correlated, which is instrumental in developing predictive models of oscillating foil arrays in which vortex wakes directly impact the energy harvesting of downstream foils.

\section{Acknowledgments}

This material is based upon work supported by the National Science Foundation under award CBET-1921594 and program director Ron Joslin. This research was conducted using computational resources and services at the Center for Computation and Visualization at Brown University. The authors thank Kenny Breuer for fruitful discussions on oscillating foil dynamics.

\bibliography{manuscript}

\end{document}